# Classical Models of the Electron Spin - Comparison of the Electric Current Model and the Magnetic Charge Model

Ferromagnetic matter finds its microscopic origin in the intrinsic electron spin, which is considered to be a purely quantum mechanical property of the electron. To incorporate the influence of the electron spin in the microscopic and macroscopic Maxwell equations - and thereby in classical physics - two models have been utilized: the electric current and the magnetic charge model. This paper aims to highlight fundamental problems of the commonly used current loop model, widely employed in textbooks. This work demonstrates that the behavior of a constant electric current dipole is not described by the laws of classical electrodynamics. More precisely, the electric current model is dependent on external forces, not included in Maxwells field and force equations, in order to maintain the force balance on the electric charge density inside the electron. These external forces change dynamically and do work on the system as the electron interacts with external fields. Consequently, the energies derived from classical physics (gravitational potential energy, kinetic energy, electrodynamic field energy) are not conserved in a system including constant electric current dipoles. In contrast to the electric current model, the magnetic charge model employs separate magnetic charges to model the electron spin, requiring the Maxwell equations to be extended by magnetic sources. This paper intends to illustrate that the magnetic charge model has significant advantages over the electric current model as it needs no external forces and energies, is a closed electromechanical system and is fully modeled by the classical laws of physics. This work forms the basis for the derivation and consideration of equivalent problems in macroscopic systems involving ferromagnetic matter.

Bela Schulte Westhoff, Technical University of Berlin (e-mail: schultewesthoff@tu-berlin.de)

## I. INTRODUCTION

Ferromagnetic materials are key components in various mechatronic systems and engineering applications, as they form the basis for converting electrical energy into mechanical movement. This includes materials such as iron, nickel, cobalt or neodymium-iron-boron. To model the behaviour of ferromagnetic materials in mechatronic systems, classical electrodynamics continues to hold its ground as an essential and highly successful theory in the realm of physics, finding widespread application in numerous fields of science and engineering. Despite being a theory not considering quantum mechanical effects, classical electrodynamics accurately describes the behavior of systems including electric charges and currents, particularly when dealing with macroscopic phenomena. In many systems, the quantum mechanical effects present are either too small to be observed or are averaged out over large scales, making it reasonable to neglect their influence. Ferromagnetic matter however has its origin in the intrinsic electron-spin, which is considered to be a purely quantum mechanical effect [1, 34-3] [1, p. 37] [2, p. 187].

The microscopic structure of solid ferromagnetic matter is illustrated schematically in Figure 1. The origin of the fields and forces of ferromagnetic matter is microscopically attributed to the magnetic dipole moment of the quantum mechanical spin property of the electron. Only about up to 1% of the macroscopic magnetic field emanating from ferromagnetic matter is attributable to the motion of matter-bound electrons, the rest is caused by the spin of electrons [3, p. 170][4, p. 408]. To circumvent the mathematical complexity of quantum physics, different classical electrodynamic models of the magnetic dipole moment of the electron-spin have been used to represent its influence in the microscopic and macroscopic Maxwell equations and thereby in classical physics. As shown in [1, 18-2] classical physics is completely described microscopically by the equations in Table 1 (as the microscopic Maxwell equations are considered in this paper, the lower-case letters for the field variables are used). Thermodynamic quantities of macroscopic systems (such as temperature or pressure) are microscopically attributable to the laws in Table 1. The energies that may be derived from the equations of classical physics in Table 1 are the gravitational potential energy, the kinetic energy and the electrodynamic field energies. In a closed electromechanical system described by the laws of classical physics, the sum of forementioned energies is conserved.

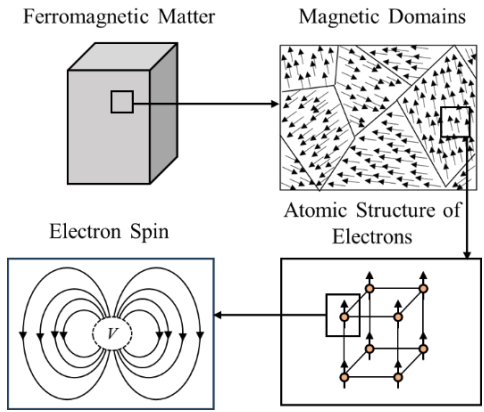

Figure 1: Schematic microscopic structure of ferromagnetic matter.

| Electrodynamics | Law of motion |
|---|---|
| Maxwells Equations: | (Newton's Law with Einstein's modification) |
| $\nabla \cdot \mathbf{e} = \dfrac{\rho_e}{\varepsilon_0}$ | $\dfrac{d}{dt}(\mathbf{p}) = \mathbf{F},$ |
| $-\nabla \times \mathbf{e} - \dfrac{\partial \mathbf{b}}{\partial t} = 0$ | $\mathbf{p}_{Newton} = m\mathbf{v}$ |
| $\nabla \cdot \mathbf{b} = 0$ | $\mathbf{p}_{Einstein} = \dfrac{m\mathbf{v}}{\sqrt{1 - v^2/c^2}}$ |
| $\nabla \times \mathbf{b} - \dfrac{1}{c^2}\dfrac{\partial \mathbf{e}}{\partial t} = \mu_0 \mathbf{j}_e$ | |
| Conservation of Charge: | Gravitation |
| $\nabla \cdot \mathbf{j} = -\dfrac{\partial \rho}{\partial t}$ | $\mathbf{F} = -G\dfrac{m_1 m_2}{r^2}\mathbf{e}_r$ |
| Force Law: | |
| $\mathbf{f}_e = \rho_e \mathbf{e} + \mathbf{j}_e \times \mathbf{b}$ | |

Table 1: Laws of Classical Physics.

The most common and generally accepted classical model of the magnetic dipole moment is the electric current model, widely employed in most textbooks on electromagnetic theory [1, 2, 4–8]. As schematically illustrated in Figure 2a), a bound electric current represents the electrons spin-behavior and models its impact in the classical field and force equations. In contrast, the magnetic charge model is used as the preferred model only by a few authors, such as [3, 9]. In the context of magnetostatic fields, the magnetic charge model is employed by other authors, such as [10, 11]. As illustrated in Figure 2b), the electrons magnetic moment is not modeled by electric sources but by separated magnetic charges of opposite signs. Accordingly, the microscopic Maxwell equations need to be extended by the magnetic charge density $\rho_\mathrm{m}$ and magnetic current density $\mathbf{j}_\mathrm{m}$ in order to describe the behavior of this model. To distinguish between systems with and without magnetic sources (and in anticipation of what will be derived), a different field variable is used to describe a system including magnetic sources in this paper. The variable of the magnetic field $\mathbf{b}$ is exchanged by the magnetic field strength $\mu_0\mathbf{h}$, which in vacuum is proportional to each other:

$$\mathbf{b} = \mu_0\mathbf{h}. \tag{1}$$

$\mu_0$ is the physics constant of the vacuum magnetic permeability. Including magnetic sources and exchanging $\mathbf{b}$ by $\mu_0\mathbf{h}$, the classical laws of physics are given by the equations in Table 2.

Different arguments have been made to favor the electric current model over the magnetic charge model. In [5, p. 242][5, p. 269] it is claimed that the magnetic charge model is "bad physics" since, to the present state of knowledge, no magnetic charges exist. In [3, p. 173] it is objected that, although no magnetic charges have yet been observed, this does not imply that they do not exist. Beyond that it is important to note that the electric current model itself is incompatible with classical physics. Given the size and field strength of the magnetic dipole, the corresponding charge velocity of the electric current would exceed the speed of light [12]. Both classical models cannot represent the entire behavior of the electron and are merely approximations of the quantum mechanical spin, as implied by [1, p. 37][2, p. 187]. Both classical models are introduced with the objective of incorporating a source term into Maxwell's field equations to represent the magnetic moment of the electron spin and allow the use of classical physics when dealing with ferromagnetic matter on a micro- and macro scale.

The aim of this paper is to illustrate that the electric current model is not a closed electrodynamic model as that the behaviour of the electron on the microscopic scale as well as the behaviour of ferromagnetic matter on a macroscopic scale cannot be described by the equations of classical physics in Table 1. Instead, the electric current model requires both external forces and energy sources, which cause the conservation of electromechanical energy in the system to be violated. In contrast, this work aims to show that the magnetic charge model of the electron spin is a closed electrodynamic system and has no need for external forces and energies. The goal is to illustrate that by extending the Maxwell equations by magnetic sources, all relevant microscopic and macroscopic behaviour in ferromagnetic matter caused by the electron spin is modelled by the laws of classical physics in Table 2.

To prove these statements, the two electron models are examined in terms of fields, forces, and power. The total force and the magnetic fields of the two dipole models have been evaluated and compared in various works [13–19]. In [18, 20] it is claimed that the total force of the models differ, allowing the possibility to experimentally rule out one of the models unambiguously. In [9, p. 834] it is stated in contrast that the forces are the same and that the behavior is indistinguishable. In this work, the origin of these contradictions is examined by considering in detail what happens within the electron volume. In the regarded literature, the electron's magnetic dipole representations are merely considered as point entities, leaving the internal dynamics within the dipole volume unaddressed.

In this work, a comprehensive examination of the phenomena occurring within the crucial electron volume illustrated in Figure 1 is conducted. The goal is to elucidate that distinguishing between electric and magnetic sources as the representation of the constant magnetic dipole is infeasible solely through observations of the fields outside the dipole volume. Notably, differences in field distributions emerge exclusively within the dipole volume, while discrepancies in the total dipole force necessitate an external source within this domain. The implications of these findings on the understanding of the classical electron spin representation will be explored, including a brief review and discussion on the hyperfine structure of atoms, which is often cited in support of the electric current model.

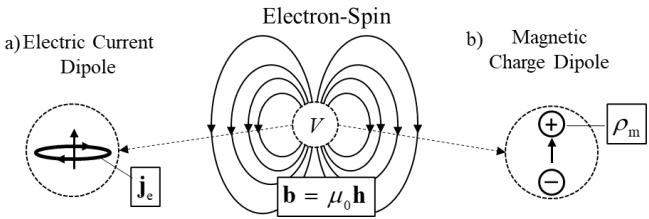

Figure 2: a) Schematic illustration of the electric current model and b) magnetic charge model of the magnetic dipole of the electron spin.

| Electrodynamics | | Law of motion |
|---|---|---|
| Maxwells Equations: | | (Newton's Law with Einstein's modification) |
| $\nabla \cdot \mathbf{e} = \dfrac{\rho_\mathrm{e}}{\varepsilon_0}$ | | |
| $-\nabla \times \mathbf{e} - \mu_0 \dfrac{\partial \mathbf{h}}{\partial t} = \mu_0 \mathbf{j}_\mathrm{m}$ | | $\dfrac{d}{dt}(\mathbf{p}) = \mathbf{F},$ |
| $\mu_0 \nabla \cdot \mathbf{h} = \mu_0 \rho_\mathrm{m}$ | | $\mathbf{p} = \dfrac{mv}{\sqrt{1 - v^2/c^2}}$ |
| $\mu_0 \nabla \times \mathbf{h} - \dfrac{1}{c^2}\dfrac{\partial \mathbf{e}}{\partial t} = \mu_0 \mathbf{j}_\mathrm{e}$ | | |
| Conservation of Charge: | | **Gravitation** |
| $\nabla \cdot \mathbf{j}_\mathrm{e} = -\dfrac{\partial \rho_\mathrm{e}}{\partial t}, \ \nabla \cdot \mathbf{j}_\mathrm{m} = -\dfrac{\partial \rho_\mathrm{m}}{\partial t}$ | | $\mathbf{F} = -G\dfrac{m_1 m_2}{r^2}\mathbf{e}_r$ |
| Force Law: | | |
| $\mathbf{f}_\mathrm{m} = \rho_\mathrm{e}\mathbf{e} + \mu_0\mathbf{j}_\mathrm{e} \times \mathbf{h} + \mu_0\rho_\mathrm{m}\mathbf{h} - \mu_0\varepsilon_0\mathbf{j}_\mathrm{m} \times \mathbf{e}$ | | |

Table 2: Laws of Classical Physics including magnetic sources.

Next to the field and total force considerations, the comparison is extended to examine the power generation and internal forces of the two models. This crucial comparison, which is rarely considered in the examined literature, highlights how the two models differ fundamentally. Here, the necessity of external forces and energies is illustrated. These external forces change dynamically and do work on the system as the electron interacts with external fields. Consequently, it is presented that the conservation of electromechanical energies in a system including constant electric dipoles is violated. The analysis concludes by illustrating that the magnetic charge model in opposition is not dependent on external sources and is entirely described by the laws of classical physics in Table 2. This paper aims to build a basis for further considerations, which include the derivation of the macroscopic electromagnetic theory based on both models, force density considerations in macroscopic systems, and force calculations using variation of energy in mechatronic systems including ferromagnetic matter.

The paper is structured in the following way: in chapter II, the extended Maxwell equations including magnetic charges and currents are presented. In chapter III, the two spin representations of the electron in terms of fields, forces, and power are compared by looking into the dipole volume in detail. This forms the foundation to argue in favor of the magnetic charge model as the representation of the electron spin. In chapter IV, it is differentiated between the electron model as a distributed and concentrated system based on its interaction with the surrounding system. Chapter V comments on a different model in which the electron spin is not associated with electric or magnetic sources but is merely considered as a constant magnetic dipole. In the final chapter a summary and discussion are given.

## II. ELECTRODYNAMICS INCLUDING MAGNETIC SOURCES

The considerations in this paper are based on the field equations of classical electromagnetism, which are briefly presented in this chapter. The established Maxwell equations without magnetic charge in vacuum are given by:

$$\begin{aligned} \nabla \cdot \mathbf{e} &= \frac{\rho_e}{\varepsilon_0} \\ \nabla \times \mathbf{e} &= -\frac{\partial \mathbf{b}}{\partial t} \\ \nabla \cdot \mathbf{b} &= 0 \\ \frac{1}{\mu_0} \nabla \times \mathbf{b} &= \mathbf{j}_e + \varepsilon_0 \frac{\partial \mathbf{e}}{\partial t} \end{aligned} \qquad (2)$$

The dynamic behavior of an electromechanic system including electric charge density $\rho_e$ and electric current density $\mathbf{j}_e$ is determined by the electric force density:

$$\mathbf{f}_e = \rho_e \mathbf{e} + \mathbf{j}_e \times \mathbf{b} \cdot \qquad (3)$$

To distinguish between electric and magnetic sources, this work uses the subscripts $_e$ and $_m$. The first term of (3) is referred to as the electric Coulomb force and the second term as the electric Lorentz force. Since the Lorentz force acts perpendicular to the motion of charge, it does not execute any power. The electromagnetic power resulting from (3) is given by:

$$P_{e,em} = \mathbf{j}_e \cdot \mathbf{e} \qquad (4)$$

Equation (3) and (4) may be reformulated in terms of fields only [2, p. 260][4, p. 507]:

$$-\nabla \cdot \underline{\underline{\mathbf{T}}}_e + \frac{\partial}{\partial t}(\mathbf{g}_e) = -\rho_e \mathbf{e} - \mathbf{j}_e \times \mathbf{b} , \qquad (5)$$

$$\nabla \cdot \mathbf{S} + \frac{\partial u_{e,em}}{\partial t} = -\mathbf{j}_e \cdot \mathbf{e} \cdot \qquad (6)$$

Here $\mathbf{T}$ denotes the Maxwell stress tensor

$$\underline{\underline{\mathbf{T}}}_e = \varepsilon_0 \left[ \mathbf{e} \otimes \mathbf{e} + c^2 \mathbf{b} \otimes \mathbf{b} - \frac{1}{2} \underline{\underline{\mathbf{I}}} \left( e^2 + c^2 b^2 \right) \right], \qquad (7)$$

$\mathbf{g}_e$ the electromagnetic linear momentum

$$\mathbf{g}_e = \varepsilon_0 \mathbf{e} \times \mathbf{b} , \qquad (8)$$

$\mathbf{S}_e$ the poynting vector (power flow) which is proportional to the linear momentum

$$\mathbf{S}_e = \frac{1}{\mu_0} \mathbf{e} \times \mathbf{b} = c^2 \mathbf{g}_e , \qquad (9)$$

and $u_{e,em}$ the electromagnetic energy density

$$u_{e,em} = \frac{\varepsilon_0}{2} \mathbf{e} \cdot \mathbf{e} + \frac{1}{2\mu_0} \mathbf{b} \cdot \mathbf{b} \cdot \qquad (10)$$

As shown in [1, p. 153], the defined field terms for the energy, linear momentum, poynting vector and stress tensor may be reformulated using vector identities. In the following, the here given definitions are used.

The duality transformation of the Maxwell equations allows all electric charges to be replaced by a sum of magnetic and electric charges without affecting any observable phenomenon, as long as their ratio is the same for each particle [4, p. 49][2, p. 273]. So far, no other ratio has been found. In other words, the field equations with solely electric charges as in (2) and (3) may be considered to be a convention. The transformation laws for the symmetrized Maxwell equations are given in [4, p. 49]. Including magnetic charges, (2) formulates to:

$$\begin{aligned} \nabla \cdot \mathbf{e} &= \frac{\rho_e}{\varepsilon_0} \\ -\nabla \times \mathbf{e} - \mu_0 \frac{\partial \mathbf{h}}{\partial t} &= \mu_0 \mathbf{j}_m \\ \mu_0 \nabla \cdot \mathbf{h} &= \mu_0 \rho_m \\ \mu_0 \nabla \times \mathbf{h} - \frac{1}{c^2} \frac{\partial \mathbf{e}}{\partial t} &= \mu_0 \mathbf{j}_e \end{aligned} \qquad (11)$$

Again, the variable $\mathbf{b} = \mu_0 \mathbf{h}$ is exchanged merely to illustrate that the system includes magnetic charges. The force equation is given by:

$$\mathbf{f}_m = \rho_e \mathbf{e} + \mu_0 \mathbf{j}_e \times \mathbf{h} + \mu_0 \rho_m \mathbf{h} - \mu_0 \varepsilon_0 \mathbf{j}_m \times \mathbf{e} \cdot \qquad (12)$$

The electromagnetic power is given by:

$$P_{m,em} = \mathbf{j}_e \cdot \mathbf{e} + \mu_0 \mathbf{j}_m \cdot \mathbf{h} \,. \qquad (13)$$

The force and power equations including magnetic charges may be reformulated in terms of fields [3, p. 275]:

$$-\nabla \cdot \underline{\underline{\mathbf{T}_m}} + \frac{\partial \mathbf{g}_m}{\partial t} = -\rho_e \mathbf{e} - \mu_0 \mathbf{j}_e \times \mathbf{h} - \mu_0 \rho_m \mathbf{h} + \mu_0 \varepsilon_0 \mathbf{j}_m \times \mathbf{e} \,, \qquad (14)$$

$$\nabla \cdot \mathbf{S}_m + \frac{\partial u_{m,em}}{\partial t} = -\mathbf{j}_e \cdot \mathbf{e} - \mu_0 \mathbf{j}_m \cdot \mathbf{h} \,. \qquad (15)$$

It is noteworthy to observe that the resultant components of the Maxwell stress tensor $\mathbf{T}_m$, the Momentum $\mathbf{g}_m$, the poynting vector $\mathbf{S}_m$ and the energy density $u_{m,em}$ are the same with and without magnetic charges [2, p. 274]:

$$\underline{\underline{\mathbf{T}_m}} = \varepsilon_0 \left[ \mathbf{e} \otimes \mathbf{e} + c^2 \mu_0^2 \mathbf{h} \otimes \mathbf{h} - \frac{1}{2} \underline{\underline{\mathbf{I}}} \left( e^2 + c^2 \mu_0^2 h^2 \right) \right], \qquad (16)$$

$$\mathbf{g}_m = \varepsilon_0 \mu_0 \mathbf{e} \times \mathbf{h} \,, \qquad (17)$$

$$\mathbf{S}_m = \mathbf{e} \times \mathbf{h} = c^2 \mathbf{g}_m \,, \qquad (18)$$

$$u_{m,em} = \frac{\varepsilon_0}{2} \mathbf{e} \cdot \mathbf{e} + \frac{\mu_0}{2} \mathbf{h} \cdot \mathbf{h} \,. \qquad (19)$$

## III. Electric and Magnetic Dipoles - Comparison of Bound Electric and Magnetic Sources in overall neutral volumes

In this chapter, electric and magnetic sources as classical representations of the electron spin are compared. The electron spin at rest exclusively evokes a magnetic field. In a moving frame however, a magnetic dipole also exhibits the behavior of an electric dipole [4, p. 859, 21, 22]. In this paper, the spins at rest or at low speeds are considered, as special relativity must be taken into account for moving electrons (strictly speaking, even at low speeds [23]). Still, this work is not restricted to examine the behavior of magnetic dipoles but also briefly considers electric dipoles. This forms a basis for considerations of the electron in moving inertia frames in future work.

In Figure 3 the spin volume in its most general form is illustrated, having the characteristics of an arbitrary electric and magnetic dipole field. This chapter is subdivided into four sections: In section *A)*, the electric and magnetic fields of electric and magnetic sources are compared, in section *B)*, the total force on the two dipole models is compared and in section C) and D) the differences in terms of internal forces, energy, and power are illustrated.

### A. Field Comparison of Electric Current Dipoles and Magnetic Charge Dipoles

The approach to compare the fields of electric and magnetic sources is the following: First, arbitrary electric sources within the volume are allowed to be the origin of the electric and magnetic fields, the only prerequisite being that they are overall neutral. Then the electric sources are exchanged by magnetic sources and it is shown that their fields merely differ within the volume. Thereby this work aims to illustrate that there is no way to determine the origin of the spin by considering the fields outside the electron volume (Figure 3).

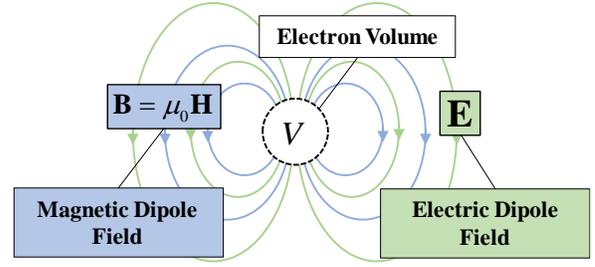

Figure 3: Overall neutral volume V originating an electric and magnetic field.

When considering the bound electric charge density $\rho_{e,b}$ and bound electric currents $\mathbf{j}_{e,b}$ in an overall neutral volume, it will be helpful to express these sources using two new vector fields (the subscript $_b$ indicates bound sources). The electric charge density $\rho_{e,b}$ shall be defined by the divergence of the vector field $\mathbf{p}$, which is called polarization field in this work:

$$\rho_{e,b} = -\nabla \cdot \mathbf{p} \,. \qquad (20)$$

In this paper surface charges and currents are neglected for mathematical simplicity, hence $\mathbf{p}$ is continuous. Since the dipole volume $V$ is overall neutral, for the electric charge density it holds:

$$\int_V \rho_{e,b} dV = 0 \,. \qquad (21)$$

Equation (21) permits $\mathbf{p}$ to be constrained to be zero outside the considered volume [4, p. 159], yielding the boundary condition (it is remarked that $\mathbf{p}$ is not uniquely defined; details in Appendix B):

$$\mathbf{p}(\mathbf{x}) = 0 \quad \mathbf{x} \notin V \,. \qquad (22)$$

When considering electric current densities as field sources in the dipole volume, two parts are distinguished. The first part results from a displacement of the charge distribution $\rho_{e,b} = -\nabla \cdot \mathbf{p}$ within the volume. Given the continuity equation

$$\frac{\partial \rho_{e,b}}{\partial t} = -\frac{\partial (\nabla \cdot \mathbf{p})}{\partial t} = -\nabla \cdot \frac{\partial \mathbf{p}}{\partial t} = -\nabla \cdot \mathbf{j}_{e,b2} \,, \qquad (23)$$

the associated electric polarization current can be derived:

$$\mathbf{j}_{e,b2} = \frac{\partial \mathbf{p}}{\partial t} \,. \qquad (24)$$

The second part is a divergence-free electric current $\mathbf{j}_{e,b}$ providing the homogeneous solution of (23):

$$\nabla \cdot \mathbf{j}_{e,b} = 0 \,. \qquad (25)$$

Again, it will be helpful to express $\mathbf{j}_{e,b}$ in terms of another vector field. As $\mathbf{j}_{e,b}$ is divergence-free, it may be expressed by the rotation of the vector field $\mathbf{m}$, which is called magnetization field in this work:

$$\mathbf{j}_{e,b} = \nabla \times \mathbf{m} \,. \qquad (26)$$

As the electric current $\mathbf{j}_{e,b}$ is restricted to the volume, $\mathbf{m}$ may be restricted to be zero everywhere outside $V$ ($\mathbf{m}$ is not uniquely defined; details in Appendix B):

$$\mathbf{m}(\mathbf{x}) = 0 \quad \mathbf{x} \notin V \, . \tag{27}$$

To sum up, arbitrary electric sources in the overall neutral dipole volume $V$ are considered, which are represented by the vector fields $\mathbf{m}$ and $\mathbf{p}$:

$$\rho_{e,b} = -\nabla \cdot \mathbf{p}, \quad \mathbf{j}_{e,b2} = \frac{\partial \mathbf{p}}{\partial t}, \quad \mathbf{j}_{e,b} = \nabla \times \mathbf{m} \, . \tag{28}$$

Given the volume is overall neutral, every possible electric source may be represented by $\mathbf{p}$ and $\mathbf{m}$, it is merely a different mathematical representation. The corresponding fields are obtained by inserting the sources into the field equations of Table 1:

$$\begin{aligned} \nabla \cdot \mathbf{e} &= -\frac{\nabla \cdot \mathbf{p}}{\varepsilon_0} \\ \nabla \times \mathbf{e} &= -\frac{\partial \mathbf{b}}{\partial t} \\ \nabla \cdot \mathbf{b} &= 0 \\ \frac{1}{\mu_0} \nabla \times \mathbf{b} &= \nabla \times \mathbf{m} + \varepsilon_0 \frac{\partial \mathbf{e}}{\partial t} + \frac{\partial \mathbf{p}}{\partial t} \end{aligned} \, . \tag{29}$$

Next, it is shown how to replace all electric sources by magnetic sources without changing the fields outside the dipole volume where $\mathbf{m}$ and $\mathbf{p}$ is zero. As the electron-spin at rest only evokes a magnetic field, its influence may be modeled exclusively by stationary electric currents and the corresponding vector field $\mathbf{m}$. To keep it simple, merely the vector field $\mathbf{m}$ is considered in the following. How to exchange the electric source corresponding to $\mathbf{p}$ by magnetic sources is shown in Appendix A. In order to exchange the sources corresponding to $\mathbf{m}$, (29) is reordered:

$$\begin{aligned} \nabla \cdot \mathbf{e} &= 0 \\ \nabla \times \mathbf{e} &= -\frac{\partial \mathbf{b}}{\partial t} \\ \nabla \cdot \mathbf{b} &= 0 \\ \frac{1}{\mu_0} \nabla \times (\mathbf{b} - \mu_0 \mathbf{m}) &= \frac{\partial \varepsilon_0 \mathbf{e}}{\partial t} \end{aligned} \, . \tag{30}$$

In foresight of what is aimed to derive, the new vector field $\mathbf{h}$ is introduced:

$$\mathbf{b} = \mu_0 (\mathbf{h} + \mathbf{m}) \quad \Leftrightarrow \quad \mathbf{h} = \frac{1}{\mu_0} \mathbf{b} - \mathbf{m} \quad . \tag{31}$$

$\mathbf{h}$ has no physical significance in the regarded system of electric sources, it is merely defined by (31). Inserting (31) into (30) yields:

$$\begin{aligned} \nabla \cdot \mathbf{e} &= 0 \\ \nabla \times \mathbf{e} &= -\mu_0 \frac{\partial \mathbf{h}}{\partial t} - \mu_0 \frac{\partial \mathbf{m}}{\partial t} \\ \nabla \cdot \mathbf{h} &= -\nabla \cdot \mathbf{m} \\ \nabla \times \mathbf{h} &= \varepsilon_0 \frac{\partial \mathbf{e}}{\partial t} \end{aligned} \, . \tag{32}$$

Comparing (32) with the magnetic field equations including magnetic sources in Table 2, it turns out that the sources generating the vector field $\mathbf{h}$ are the magnetic charge density $\rho_{m,b}$ and the magnetic current densities $\mathbf{j}_{m,b}$:

$$\mathbf{j}_{m,b} = \frac{\partial \mathbf{m}}{\partial t}, \quad \rho_{m,b} = -\nabla \cdot \mathbf{m} \, . \tag{33}$$

In Figure 4 an example of a stationary vector field $\mathbf{m}$ and the corresponding electric and magnetic sources are schematically illustrated. In the system with the electric sources $\mathbf{j}_{e,b} = \nabla \times \mathbf{m}$, the relevant physical field is given by $\mathbf{b}$ while the field $\mathbf{h}$ (as defined by (31)) has no physical significance. In the system with magnetic sources $\mathbf{j}_{m,b} = \frac{\partial \mathbf{m}}{\partial t}$ and $\rho_{m,b} = -\nabla \cdot \mathbf{m}$, the relevant physical field is given by $\mu_0 \mathbf{h}$ while the field $\mathbf{b}$ (as defined by (31)) has no physical significance. As sown by (31), the fields $\mathbf{b}$ and $\mu_0 \mathbf{h}$ solely differ inside the considered volume by $\mu_0 \mathbf{m}$. Following, the electric current density $\mathbf{j}_{e,b}$ generates the same field outside the dipole volume as the magnetic charges density $\rho_{m,b}$ and the magnetic current density $\mathbf{j}_{m,b}$. To distinguish magnetic and electric sources (or a superposition) as origin of the spin fields by observing the fields outside the electron volume is not possible.

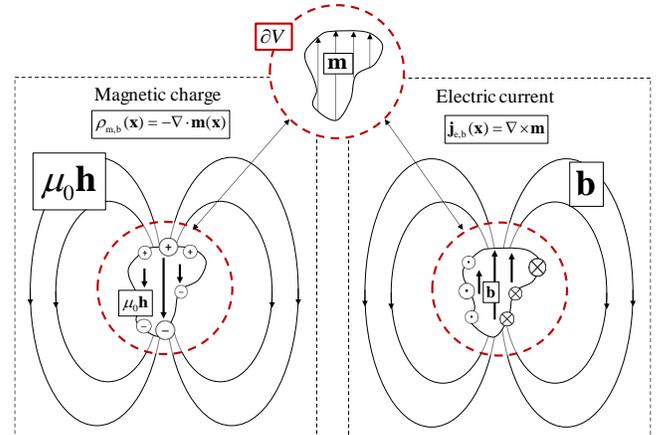

Figure 4: a) Magnetic charge representation of an arbitrary vector field m b) Electric current representation of an arbitrary vector field m.

This section is concluded with a brief commentary and outlook on the findings. The vector fields $\mathbf{m}$ and $\mathbf{p}$ were introduced for no other reason but to represent electric or magnetic sources in an overall neutral volume. It was illustrated, how to microscopically interpret the fields $\mathbf{b}$ and $\mu_0 \mathbf{h}$ and assign them physical origins. Everything demonstrated in this section may be transferred to the macroscopic Maxwell equations. The transition from microscopic to macroscopic Maxwell equations is merely a rearrangement of sources and corresponding fields of a system using averaging functions (as will be shown in future work). In many textbooks, the macroscopic magnetic field strength $\mathbf{H}$ and the displacement field $\mathbf{D}$ (Appendix A) are introduced, without relating them to any meaningful physical interpretation, but just for mathematical convenience. Surely, this is reasonable in systems without magnetic sources, as they have no physical interpretation in these systems. However, even on a macroscopic scale, the magnetic field strength $\mathbf{H}$ may be regarded to be the origin of magnetic sources and $\mathbf{B}$ to be the origin of electric sources corresponding to the vector field $\mathbf{M}$.

### B. Total Force Comparison of Electric Current Dipoles and Magnetic Charge Dipoles

In section III A it was shown that the origin of the electron spin cannot be identified by considering the fields outside the electron volume. In this section, the total force on a vector field $\mathbf{m}$ and its corresponding electric and magnetic sources is compared (the same may be derived for $\mathbf{p}$):

$$\mathbf{j}_{e,b} = \nabla \times \mathbf{m} \quad \leftrightarrow \quad \mathbf{j}_{m,b} = \frac{\partial \mathbf{m}}{\partial t}, \rho_{m,b} = -\nabla \cdot \mathbf{m}. \tag{34}$$

More precisely, not the total force is compared, but the change of the mechanical momentum of the electron's center of mass, as this can be observed experimentally. Therefore, the dipole is assigned the rest mass $m_r$ and the moment of inertia $J_r$. First, the electric current dipole is considered. The electromagnetic force $\mathbf{F}_{e,em}$ caused by an external magnetic field $\mathbf{b}_{ext}$ equals the change of the total mechanical momentum $\mathbf{P}_{e,mech}$ of the volume:

$$\frac{d\mathbf{P}_{e,mech}}{dt} = \mathbf{F}_{e,em} = \int \mathbf{j}_{e,b} \times \mathbf{b}_{ext} \, dV = \int (\nabla \times \mathbf{m}) \times \mathbf{b}_{ext} \, dV. \tag{35}$$

Besides the linear moment of the center of mass $\mathbf{P}_{e,cm}$, a hidden mechanical momentum $\mathbf{P}_{e,hid}$ may be present in the dipole. More details on the hidden momentum in dipoles are given in [4, p. 521] [14, 24–26]. Following, $\mathbf{P}_{e,mech}$ is composed of:

$$\mathbf{P}_{e,mech} = \mathbf{P}_{e,cm} + \mathbf{P}_{e,hid}. \tag{36}$$

The observable change of momentum of the center of mass results in:

$$\frac{d\mathbf{P}_{e,cm}}{dt} = \frac{d\mathbf{P}_{e,mech}}{dt} - \frac{d\mathbf{P}_{e,hid}}{dt} = \int (\nabla \times \mathbf{m}) \times \mathbf{b}_{ext} \, dV - \frac{d\mathbf{P}_{e,hid}}{dt}. \tag{37}$$

As presented in [4, p. 521] [14, 24–26], the hidden moment $\mathbf{P}_{e,hid}$ of the considered volume is balanced by a change of linear electromagnetic moment. This follows from the general theorem that the total momentum of a closed system is zero when its center of energy is at rest. The linear electromagnetic moment of the volume $\mathbf{P}_{e,em}$ gets induced by an external electric field $\mathbf{e}_{ext}$ and equals ( [4, p. 521] [14, 24–26]):

$$-\mathbf{P}_{e,hid} = \mathbf{P}_{e,em} = \int \varepsilon_0 \mathbf{e}_{ext} \times \mathbf{b} \, dV. \tag{38}$$

Inserting (38) in (37), it follows:

$$\frac{d\mathbf{P}_{e,cm}}{dt} = \int (\nabla \times \mathbf{m}) \times \mathbf{b}_{ext} \, dV + \frac{d\int \varepsilon_0 \mathbf{e}_{ext} \times \mathbf{b} \, dV}{dt}. \tag{39}$$

For small non-relativistic velocities $\mathbf{v}_e$ of the electron it follows:

$$m_r \frac{d\mathbf{v}_e}{dt} = \int (\nabla \times \mathbf{m}) \times \mathbf{b}_{ext} \, dV + \frac{d\int \varepsilon_0 \mathbf{e}_{ext} \times \mathbf{b} \, dV}{dt}. \tag{40}$$

Next, the change of momentum of the dipole consisting of magnetic sources (34) is considered. Following the equations in Table 2 and considering the hidden momentum it follows:

$$m_r \frac{d\mathbf{v}_m}{dt} = \int \left( \mu_0 \rho_{m,b} \mathbf{h}_{ext} - \mu_0 \varepsilon_0 \mathbf{j}_{m,b} \times \mathbf{e}_{ext} \right) dV - \frac{d\mathbf{P}_{m,hid}}{dt}$$
$$= \int \mu_0 (\nabla \cdot \mathbf{m}) \mathbf{h}_{ext} - \frac{1}{c^2} \frac{\partial \mathbf{m}}{\partial t} \times \mathbf{e}_{ext} \, dV \quad . \tag{41}$$
$$+ \frac{d\int \varepsilon_0 \mu_0 \mathbf{e}_{ext} \times \mathbf{h} \, dV}{dt}$$

To compare (40) and (41), (41) is reformulated using vector identities. The first term of (41) equals [4, p. 437]:

$$\int (\nabla \cdot \mathbf{m}) \cdot \mathbf{h}_{ext} \, dV = -\int (\mathbf{m} \cdot \nabla) \mathbf{h}_{ext} \, dV + \oiint (\mathbf{n} \cdot \mathbf{m}) \mathbf{h}_{ext} \, dA. \tag{42}$$

Since $\mathbf{m}$ is zero outside the volume, the surface integral can be neglected. Following [4, p. 437] (42) equals:

$$-\int (\mathbf{m} \cdot \nabla) \mathbf{h}_{ext} \, dV = -\int \left( \mathbf{m}(\nabla \cdot \mathbf{h}_{ext}) + (\nabla \times \mathbf{m}) \times \mathbf{h}_{ext} \right)$$
$$+ (\nabla \times \mathbf{h}_{ext}) \times \mathbf{m} \right) dV - \oiint (\mathbf{m} \times \mathbf{n}) \mathbf{h}_{ext} \, dA. \tag{43}$$

Again, the surface integral equals zero. If there are no external magnetic sources within the dipole volume where $\mathbf{m}(\mathbf{x})$ is nonzero, it holds that:

$$-\int \left( \mathbf{m}(\nabla \cdot \mathbf{h}_{ext}) \right) dV = 0. \tag{44}$$

To sum up, the change of momentum is given by:

$$m_r \frac{d\mathbf{v}_m}{dt} = \int \mu_0 (\nabla \times \mathbf{m}) \times \mathbf{h}_{ext} + \mu_0 (\nabla \times \mathbf{h}_{ext}) \times \mathbf{m}$$
$$- \frac{1}{c^2} \frac{\partial \mathbf{m}}{\partial t} \times \mathbf{e}_{ext} \, dV + \frac{d\int \varepsilon_0 \mu_0 \mathbf{e}_{ext} \times \mathbf{h} \, dV}{dt}. \tag{45}$$

To compare (45) and (40), the relationship of their fields is used, which is given by:

$$\mathbf{b} = \mu_0 \left( \mathbf{h} + \mathbf{m} \right) \quad \Leftrightarrow \quad \mu_0 \mathbf{h} = \mathbf{b} - \mu_0 \mathbf{m}. \tag{46}$$

For the external fields, the relationship is given by:

$$\mathbf{b}_{ext} = \mu_0 \mathbf{h}_{ext}. \tag{47}$$

Inserting in (45), results in:

$$m_r \frac{d\mathbf{v}_m}{dt} = \int (\nabla \times \mathbf{m}) \times \mathbf{b}_{ext} + (\nabla \times \mathbf{b}_{ext}) \times \mathbf{m} - \frac{1}{c^2} \frac{\partial \mathbf{m}}{\partial t} \times \mathbf{e}_{ext} \, dV$$
$$+ \frac{d\int \varepsilon_0 \mathbf{e}_{ext} \times \mathbf{b} \, dV}{dt} + \frac{d\int \mu_0 \varepsilon_0 \mathbf{m} \times \mathbf{e}_{ext} \, dV}{dt}. \tag{48}$$

Calculating the difference between the two systems (48) and (40) yields:

$$m_r \frac{d\left( \mathbf{v}_m - \mathbf{v}_e \right)}{dt} = \int (\nabla \times \mathbf{b}_{ext}) \times \mathbf{m} \, dV$$
$$- \int \frac{1}{c^2} \frac{\partial \mathbf{m}}{\partial t} \times \mathbf{e}_{ext} \, dV + \frac{d\int \mu_0 \varepsilon_0 \mathbf{m} \times \mathbf{e}_{ext} \, dV}{dt}. \tag{49}$$
$$= \int (\nabla \times \mathbf{b}_{ext}) \times \mathbf{m} \, dV + \mu_0 \varepsilon_0 \int \mathbf{m} \times \frac{\partial \mathbf{e}_{ext}}{\partial t} \, dV$$

Inserting the Maxwell equation

$$\nabla \times \mathbf{b}_{ext} = \mu_0 \mathbf{j}_{e,ext} + \frac{1}{c^2} \frac{\partial \mathbf{e}_{ext}}{\partial t}, \tag{50}$$

(49) results in:

$$m_r \frac{d\left( \mathbf{v}_m - \mathbf{v}_e \right)}{dt} = \int \mu_0 \mathbf{j}_{e,ext} \times \mathbf{m} \, dV. \tag{51}$$

If no external electric current $\mathbf{j}_{e,ext}$ is within the dipole volume, where $\mathbf{m}$ is non-zero, the change of linear momentum is the same for both models.

In summary, it was shown that the models are indistinguishable if there are no external sources of magnetic fields ($\rho_{m,ext}$ or $\mathbf{j}_{e,ext}$) within the dipole volume. This is the expected result, as the fields $\mathbf{b}$ and $\mu_0\mathbf{h}$ differ inside the volume and consequently the force on external sources within these fields as well. Since the fields outside the volume are the same, the principle of action and reaction with external systems makes the indistinguishability of the two dipole sources evident.

This result may be emphasized by a different approach to compare the change of linear momentum of the center of mass of the dipole models. For a system including magnetic sources, it holds:

$$-\mathbf{f}_m = -\rho_e\mathbf{e}_{ext} - \mu_0\mathbf{j}_e \times \mathbf{h}_{ext} - \mu_0\rho_m\mathbf{h}_{ext} + \mu_0\varepsilon_0\mathbf{j}_m \times \mathbf{e}_{ext}$$
$$= -\nabla \cdot \underline{\underline{\mathbf{T}_m}} + \frac{\partial}{\partial t}(\mu_0\varepsilon_0\mathbf{e}_{ext} \times \mathbf{h}) \qquad (52)$$

As shown in (41), the change of momentum of the center of mass equals:

$$m_r\frac{d\mathbf{v}}{dt} = \int -\mathbf{f}_m - \frac{\partial}{\partial t}(\varepsilon_0\mu_0\mathbf{e}_{ext} \times \mathbf{h}) \, dV = \int -\nabla \cdot \underline{\underline{\mathbf{T}_m}} \, dV$$
$$= \oiint_A \underline{\underline{\mathbf{T}_m}} \cdot d\mathbf{A} \qquad (53)$$

The same follows for the electric current model. The change of the velocity of the center of mass of the dipoles is therefore only dependent on the Maxwell stress tensor. As the magnetic fields at the border of the electron volume are the same, also the superposition with external fields and thus all components of the Maxwell stress tensors $\underline{\underline{\mathbf{T}_m}}$ and $\underline{\underline{\mathbf{T}_e}}$ are identical as well. The same argument may be extended to the change of angular momentum:

$$\frac{d\mathbf{L}_{m,mech}}{dt} = \int \mathbf{r} \times (\rho_e\mathbf{e}_{ext} + \mu_0\mathbf{j}_e \times \mathbf{h}_{ext}$$
$$+ \mu_0\rho_m\mathbf{h}_{ext} - \mu_0\varepsilon_0\mathbf{j}_m \times \mathbf{e}_{ext})dV \qquad (54)$$

which equals:

$$\frac{d\mathbf{L}_{m,mech}}{dt} = -\int \left(\nabla \cdot \left(\underline{\underline{\mathbf{T}_m}} \times \mathbf{r}\right) + \frac{\partial}{\partial t}\mathbf{r} \times (\varepsilon_0\mu_0\mathbf{e}_{ext} \times \mathbf{h})\right)dV \qquad (55)$$

Considering the hidden rotational momentum $\mathbf{L}_{m,hid}$ it follows:

$$J_r\frac{d\omega}{dt} = \frac{d\left(\mathbf{L}_{m,mech} - \mathbf{L}_{m,hid}\right)}{dt}$$
$$= \frac{d\mathbf{L}_{m,mech}}{dt} + \frac{\partial}{\partial t}\mathbf{r} \times (\varepsilon_0\mu_0\mathbf{e}_{ext} \times \mathbf{h}) \qquad (56)$$
$$= -\int \nabla \cdot \left(\underline{\underline{\mathbf{T}_m}} \times \mathbf{r}\right)dV$$

The same follows for the electric current model, thus both models equal in change of angular momentum. To sum up, it was shown that the only way to distinguish the dipole models is by interaction of external sources inside the electron volume, where $\mathbf{m} \neq 0$.

At this point hyperfine splitting is discussed, a phenomenon defined by small shifts in energy levels resulting from electromagnetic interaction between the magnetic moments arising from spin of both the nucleus and electrons in atoms. The hyperfine structure unequivocally favors the electric current model for different spin particles [27]. A well-known example is the hyperfine splitting of hydrogen, which results from the interaction between the proton's and the electron's dipole field in the atom [4, p. 419] [28, 29]. To calculate the energy shift, it is necessary to depart from a purely classical electrodynamic to a partly quantum mechanical representation of the particle spin and assign its location a probability distribution. The magnitude of the energy shift unequivocally favors the electric current model for the proton spin of the hydrogen atom. Merely the fact that the electron resides inside the larger proton with a certain probability allows to distinguish the classical dipole models of the proton [28, 30]. This underlines the previous results that differentiate between the classical dipole models is only possible when further sources are inside the dipole. In [27] an overview on hyperfine structure of further atoms are presented. As the electron is such a small particle merely the interaction with a particle of similar or smaller size would allow an interaction with its inner field. In [27] it stated that the interaction between the positron's and the electron's dipole field in the positronium atom favors the electric current model for the electron. However, it needs an exotic atom like positronium and a virtual annihilation of the electron and positron (process in quantum field theory) in order to detect any differences between the models. (Side note: in my opinion it is not clearly stated in [27] how it can be concluded that the electric current model is assigned to the electron and not to the positron, since both are of the same size.)

Notwithstanding this, it is recalled that in this work it is aimed to model solid ferromagnetic matter for engineering applications in macroscopic mechatronic systems, in which the quantum mechanical effects present are either too small to be observed or are averaged out over large scales. In solid ferromagnetic matter, electrons are arranged in a crystal lattice and their distance is several orders of magnitude greater than their size. In chapter IV it will be illustrated in more detail how the lattice arrangement of solid ferromagnetic matter implies that no electron resides inside another electron. Moreover, it is recalled that both classical models cannot represent the entire behavior of the electron and are merely approximations of the quantum mechanical spin. Hyper splitting cannot be explained using purely classical electromagnetic equations, independent of the model used as quantum field theory is needed to model its effects. In the following section, it is demonstrated why the magnetic charge model, as a purely electrodynamic model, advantages over the electric current model.

### C. Internal Forces, Energy and Power of Non-Moving Electric Current and Magnetic Charge Dipoles

In the last section, the total force on the different dipole models of the electron-spin was compared. So far, no way to differentiate between a dipole consisting of electric or magnetic sources was found, given the prerequisite that nothing resides in the electron volume. In this section, internal forces, energies, and power of the dipoles are compared and it is presented in which way the two models essentially differ. It is quite evident that the electromagnetic power of the two models is different:

$$P_{\text{em,m}} = \mu_0 \mathbf{j}_{\text{m}} \cdot \mathbf{h}_{\text{ext}} \quad \leftrightarrow \quad P_{\text{em,e}} = \mathbf{j}_{\text{e}} \cdot \mathbf{e}_{\text{ext}} \cdot \tag{57}$$

The magnetic current of the magnetic charge model generates power in an external magnetic field $\mathbf{h}_{\text{ext}}$ while the electric current of the electric current model generates power in an external electric field $\mathbf{e}_{\text{ext}}$. To compare the dipole models, it must first be defined how the charges and currents are bound in the volume. For the magnetic charge dipole illustrated in Figure 5a), it is imposed that the magnetic charge distribution in the volume is rigid, meaning the magnetization in the dipole reference frame shall be constant:

$$\frac{d\mathbf{m}}{dt} = 0 \cdot \tag{58}$$

This imposes a restriction on the magnetic charges in the volume, which was not made in the previous chapters. All charges are assumed to be hold in their position by external non-electromagnetic stresses. This is comparable to the charged sphere model of an electron which is hold together by external stresses, as pointed out by Poincaré [4, p. 510, 31] (Poincaré-stresses). Consequently, no volume-bound magnetic current can arise in this non-moving dipole model. The power generated by the non-moving dipole in an arbitrary external magnetic field $\mathbf{h}$ is thus zero:

$$\mathbf{P}_{\text{em,m}} = \int \mathbf{j}_{\text{m,b}} \cdot \mathbf{h}_{\text{ext}} \, dV = 0 \cdot \tag{59}$$

A magnetic current and associated power generation can only occur as a result of translation or rotation of the electron volume, but not within the volume itself. The energy of a stationary magnetic charge dipole is given by (19):

$$U_{\text{m,dp}} = \int \frac{\mu_0}{2} \mathbf{h} \cdot \mathbf{h} dV \cdot \tag{60}$$

(60) may be reformulated in terms of $\mathbf{m}$ (Appendix A):

$$U_{\text{m,dp}} = -\int \frac{\mu_0}{2} \mathbf{m} \cdot \mathbf{h} dV \cdot \tag{61}$$

As $\mathbf{m}$ and $\mathbf{h}$ point in opposite directions within the dipole, (61) includes a minus sign.

Next, the behavior of an arbitrary shaped electric current dipole is considered, as exemplified in Figure 5b). Again, first it is considered how the currents are bound in the electron volume. The electric current model may be compared to an electric current in a neutral superconducting wire. As the wire is neutral, the electric current composes of positive and negative electric charge densities $\rho_{\text{e+}}$ and $\rho_{\text{e-}}$ which neutralize each other:

$$\rho_{\text{e+}} = -\rho_{\text{e-}}. \tag{62}$$

Their corresponding velocities $\mathbf{v}_{+}$ and $\mathbf{v}_{-}$ have the same direction $\mathbf{e}_{\text{v}}$ but their magnitudes $v_{+}$ and $v_{-}$ differ. The difference of the velocities is given by:

$$\Delta \mathbf{v} = v_{+} \mathbf{e}_{\text{v}} - v_{-} \mathbf{e}_{\text{v}} = \Delta v \mathbf{e}_{\text{v}}. \tag{63}$$

The resultant current density is electrically neutral and proportional to $\Delta v$:

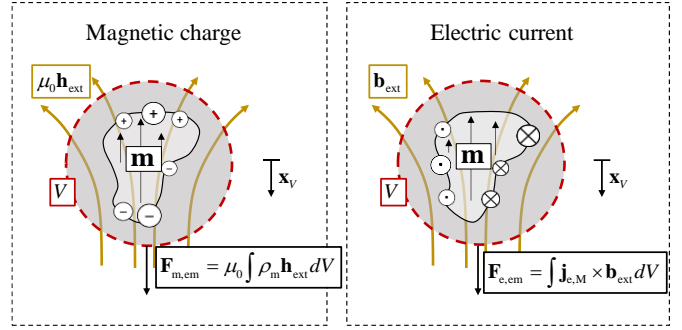

Figure 5: a) Magnetic charge representation of an arbitrary vector field m in an external magnetic field b) electric current representation of an arbitrary vector field m in an external magnetic field.

$$\mathbf{j}_{\text{e,M}} = \mathbf{j}_{\text{e,M+}} + \mathbf{j}_{\text{e,M-}} = \rho_{\text{e+}} \mathbf{v}_{+} + \rho_{\text{e-}} \mathbf{v}_{-} = \rho_{\text{e+}} \Delta \mathbf{v} = -\rho_{\text{e-}} \Delta \mathbf{v} \cdot \tag{64}$$

The electric current density of the dipole

$$\mathbf{j}_{\text{e,b}} = \rho_{\text{e+}} \Delta v \mathbf{e}_{\text{v}} = \nabla \times \mathbf{m} \tag{65}$$

is restricted to solely flow in the assigned direction of $\nabla \times \mathbf{m}$, even when interacting with external fields. The other degrees of freedom are counterbalanced by external non-electromagnetic stresses (equivalent to Poincaré Stresses). The current has one degree of freedom, comparable to an electric current in a neutral, superconducting wire.

The electromagnetic field energy of the dipole is given by (10):

$$U_{\text{e,dp}} = \int \frac{1}{2\mu_0} \mathbf{b} \cdot \mathbf{b} dV \cdot \tag{66}$$

This may be reformulated in terms of $\mathbf{m}$ (Appendix B):

$$U_{\text{e,dp}} = \int \frac{1}{2} \mathbf{m} \cdot \mathbf{b} dV \cdot \tag{67}$$

It is worth noting the difference in the sign between (67) and (60).

Next, the dipoles behavior in external fields is considered, which is comparable to that of a superconductor in an external field (the inductive behavior in external fields is presented in textbooks as [1, p. 163][4, p. 371]). As shown by (57), external electric fields $\mathbf{e}_{\text{ext}}$ are needed to generate power. The total force on the dipole volume caused by an external electric field is zero, as the forces on the charge densities $\rho_{\text{e+}}$ and $\rho_{\text{e-}}$ from external electric fields point in opposite directions and cancel each other:

$$\mathbf{f}_{\text{e,+,ext}} = \rho_{\text{e+}} \mathbf{e}_{\text{ext}} = -\rho_{\text{e-}} \mathbf{e}_{\text{ext}} = -\mathbf{f}_{\text{e,-,ext}} \cdot \tag{68}$$

However, $\mathbf{f}_{\text{e,+,ext}}$ and $\mathbf{f}_{\text{e,-,ext}}$ have an influence on the internal dipole behavior. The external electric field may be separated into its solenoidal and irrotational part. Merely solenoidal external electric fields generate a total internal force on $\rho_{\text{e+}}$ as well as on $\rho_{\text{e-}}$ around a closed current loop, as the nonrotational fields cancel out. An external solenoidal electric field is induced by a changing external magnetic field:

$$\nabla \times \mathbf{e}_{\text{ext}} = -\frac{\partial \mathbf{b}_{\text{ext}}}{\partial t} \cdot \tag{69}$$

If the inertia of the charge densities is neglected, which is assumed here, the sum of forces in direction of current flow must always be zero, for both the negative and the positive charge densities. Following, the forces from external electric fields, which point in opposite directions for the positive and negative charge densities, must be counterbalanced by further forces $\mathbf{f}_{e,+,int}$ and $\mathbf{f}_{e,-,int}$ :

$$\sum \mathbf{f}_{\rho_+} = 0 = \rho_+ \mathbf{e}_{ext} + \mathbf{f}_{e,+,int} \,,$$
$$\sum \mathbf{f}_{\rho_-} = 0 = \rho_- \mathbf{e}_{ext} + \mathbf{f}_{e,-,int} \tag{70}$$

In a superconductor, these forces are generated by an induced intrinsic electric field $\mathbf{e}_{int}$ :

$$\sum \mathbf{f}_{\rho_+} = 0 = \rho_+ \mathbf{e}_{ext} + \mathbf{f}_{e,+,int} = \rho_+ \mathbf{e}_{ext} + \rho_+ \mathbf{e}_{int} = \rho_+ \mathbf{e} \,,$$
$$\sum \mathbf{f}_{\rho_-} = 0 = \rho_- \mathbf{e}_{ext} + \mathbf{f}_{e,-,int} = \rho_- \mathbf{e}_{ext} + \rho_- \mathbf{e}_{int} = \rho_- \mathbf{e} \tag{71}$$

This intrinsic electric field is induced by a change of the electric current in the volume itself. As the forces $\rho_+ \mathbf{e}_{ext}$ and $\rho_- \mathbf{e}_{ext}$ point in opposite directions, the velocities $v_+$ and $v_-$ of the different charge densities $\rho_{e+}$ and $\rho_{e-}$ change opposite in sign, thereby changing the current density of the dipole and the corresponding internal magnetic field $\mathbf{b}_{int}$ . This change induces the compensating intrinsic electric field:

$$\nabla \times \mathbf{e}_{int} = -\frac{\partial \mathbf{b}_{int}}{\partial t} \,. \tag{72}$$

The described behavior of an electric current loop is schematically illustrated in Figure 6, while the same insights hold for arbitrary current distributions.

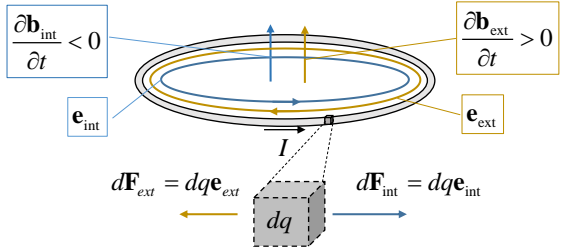

Figure 6: Schematic illustration of the behavior of a superconducting wire in an external time changing magnetic field.

Due to the inductive effect, the electric current $\mathbf{j}_{e,b}$ and the corresponding magnetizing field $\mathbf{m}$ do not remain constant in an external, time-varying magnetic field. Instead, due to the equality of forces, the total rotating electric field is zero and thus the total magnetic field in the dipole volume remains constant:

$$\nabla \times (\mathbf{e}_{ext} + \mathbf{e}_{int}) = -\frac{\partial (\mathbf{b}_{ext} + \mathbf{b}_{int})}{\partial t} = 0 \,. \tag{73}$$

The same behavior is to be observed in a superconductor, as no field enters the material and the total magnetic flux stays constant.

The described behavior fundamentally distinguishes the electric current and magnetic charge dipole. While the magnetic charge dipole stays constant, even in a time varying external magnetic field, the current density of the electric current dipole changes. However, this is not the behavior that is aimed to be modeled for the electron spin, as the dipole is a constant property of the electron. It was illustrated, that the superconducting electric current model does not represent the behavior of the electron. To keep the electric current dipole constant, the model needs to be extended by external forces, which compensate the forces resulting from the externally induced electric field $\mathbf{e}_{ext}$ , obviating the need for the self-induced electric field $\mathbf{e}_{int}$ :

$$\mathbf{f}_{\rho_+,ext} = -\rho_+ \mathbf{e}_{ext} \,,$$
$$\mathbf{f}_{\rho_-,ext} = -\rho_- \mathbf{e}_{ext} \tag{74}$$

These external forces $\mathbf{f}_{\rho_+,ext}$ and $\mathbf{f}_{\rho_-,ext}$ act on $\rho_+$ and $\rho_-$ respectively, point in opposite directions and are very different from the self-induced forces $\mathbf{f}_{e,+,int}$ and $\mathbf{f}_{e,-,int}$ in (71). These external forces are not caused by an electric field und thus neither electrodynamic nor described by the laws of classical physics given in Table 1. Also, they are very different to the before mentioned Poincaré stresses, as they dynamically change and even do work on the system. Whenever the external magnetic flux through the electron volume changes, these external forces need to be "activated" and do electromagnetic work on the system.

In order to illustrate how strange these external forces as well as a closed loop with a constant electric current are, one may think about a question in the following scenario: an arbitrary superconducting wire shall be given which forms a closed loop. Is it possible to create an electric current in the loop and keep it constant, even if external magnetic field enters? To my knowledge, this is not possible. The scenario is very different from that of an open loop, or an electric coil in which the current may be kept constant by generating different electric potentials at the ends of the wire using an ideal generator. In a closed loop, there are no two different points (two ends of a wire), where a potential difference may be applied externally. (The question may also be formulated as: How can you change the total magnetic flux through a superconducting closed loop?)

Moreover, the external forces do work on $\rho_{e+}$ and $\rho_{e-}$, which corresponds to the power:

$$\begin{aligned} P_{ext} &= \mathbf{f}_{\rho_+,ext} \cdot \mathbf{v}_+ + \mathbf{f}_{\rho_-,ext} \cdot \mathbf{v}_- \\ &= -(\rho_{e-} \mathbf{e}_{ext} \cdot \mathbf{v}_+ - \rho_{e+} \mathbf{e}_{ext} \cdot \mathbf{v}_-) = \int -\mathbf{j}_{e,b} \cdot \mathbf{e}_{ext} \, dV \\ &= \int -(\nabla \times \mathbf{m}) \cdot \mathbf{e}_{ext} \, dV \\ &= \int -\mathbf{m} \cdot (\nabla \times \mathbf{e}_{ext}) \, dV - \int \nabla \cdot (\mathbf{e}_{ext} \times \mathbf{m}) \, dV \\ &= -\int \mathbf{m} \cdot \frac{\partial \mathbf{b}_{ext}}{\partial t} \, dV \end{aligned} \tag{75}$$

This corresponds to an exchange of external energy:

$$U_{ext} = -\int\int \mathbf{j}_{e,b} \cdot \mathbf{e}_{ext} \, dV dt = -\int \int_0^{\mathbf{b}_{ext}} \mathbf{m} \cdot d\mathbf{b}_{ext} \, dV = -\int \mathbf{m} \cdot \mathbf{b}_{ext} \, dV \,. \tag{76}$$

To sum up, it was shown that the model of a constant electric current dipole as a classical model for the electron spin is not a closed electromagnetic system, as external energy as well as external forces need to be brought in the system. This statement may be generalized even further as that a constant electric current dipole is not described by the laws of classical physics given in Table 1. In contrast, no external force or power is needed to keep the magnetic charge dipole constant. The constant magnetic charge dipole is described by the laws of classical physics in Table 2.

### D. Internal Forces, Energy and Power of Two Interacting Dipoles

To further illustrate the consequeneces of including external forces and energies into the electric current dipole model of the electron, the interaction of two dipoles of magnetic and electric sources is considered in this section, as illustrated in Figure 7. (It shall be noted that relativistic effects must be taken into account for the accurate consideration of moving dipoles, even at low velocities. This will be considered in future work.)

First, the system of the magnetic charge dipole is considered (Figure 7a)). When moved, no power is needed to keep the magnetic charge distributions within the volumes constant. The volume bound magnetic current, and thus the internal power is always zero:

$$\mathbf{j}_{m,bound} = \frac{d\mathbf{m}_1}{dt} = \frac{d\mathbf{m}_2}{dt} = 0 \rightarrow P_{bound} = \mathbf{j}_{m,bound} \cdot \mathbf{h} = 0 \cdot \quad (77)$$

As one dipole is moved in the direction of force, mechanical work is done and the electromagnetic energy of the system

$$U_{m,em} = \int \frac{\mu_0}{2} \mathbf{h} \cdot \mathbf{h} dV = \int -\frac{1}{2} \mathbf{m} \cdot \mathbf{h} dV \quad (78)$$

decreases. The total electromechanical energy of the system is conserved. The force, on dipole 2 for example, may be calculated from the conservation of energy:

$$\mathbf{F}_{m,V2} = -\frac{\partial U_{em,m}}{\partial \mathbf{x}_{V2}} = -\frac{\partial \int \frac{\mu_0}{2} \mathbf{h} \cdot \mathbf{h} dV}{\partial \mathbf{x}_{V2}}. \quad (79)$$

The total field energy shall be considered in more detail. The fields are separated into the fields of the individual dipoles:

$$\mathbf{h} = \mathbf{h}_1 + \mathbf{h}_2 \qquad \mathbf{m} = \mathbf{m}_1 + \mathbf{m}_2 \cdot \quad (80)$$

The total energy is divided into the mutual and own energy terms of the two dipoles:

$$U_{m,em} = \int \frac{\mu_0}{2} \mathbf{h}_{V1} \cdot \mathbf{h}_{V1} dV + \int \frac{\mu_0}{2} \mathbf{h}_{V2} \cdot \mathbf{h}_{V2} dV \\ + \int \mu_0 \mathbf{h}_{V1} \cdot \mathbf{h}_{V2} dV \quad (81)$$

Since the magnetic charge distributions of each dipole remain constant when moved, the magnetic fields and energy terms of each dipole remain constant:

$$\frac{\partial \left( \int \frac{\mu_0}{2} \mathbf{h}_{V1} \cdot \mathbf{h}_{V1} dV \right)}{\partial \mathbf{x}_{V2}} = \frac{\partial \left( \int \frac{\mu_0}{2} \mathbf{h}_{V2} \cdot \mathbf{h}_{V2} dV \right)}{\partial \mathbf{x}_{V2}} = 0 \cdot \quad (82)$$

As work is done, merely the mutual term of the electromagnetic energy (81) of the system decreases. The total force on dipole 2 (illustrated in Figure 7) equals the change in mutual electromagnetic energy:

$$\mathbf{F}_{m,V2} = \int \mu_0 \rho_{m,V2} \mathbf{h} dV = -\frac{\partial U_{em,m}}{\partial \mathbf{x}_{V2}} = -\frac{\partial \left( \int \mu_0 \mathbf{h}_{V1} \cdot \mathbf{h}_{V2} dV \right)}{\partial \mathbf{x}_{V2}}. \quad (83)$$

As shown in Appendix, (83) may be reformulated in terms of $\mathbf{m}$:

$$\mathbf{F}_{m,V2} = -\frac{\partial \left( -\int \mu_0 \mathbf{m}_{V1} \cdot \mathbf{h}_{V2} dV \right)}{\partial \mathbf{x}_{V2}} = \frac{\partial \left( \int \mu_0 \mathbf{m}_{V2} \cdot \mathbf{h}_{V1} dV \right)}{\partial \mathbf{x}_{V2}}. \quad (84)$$

The corresponding magnetic power is attributable to the current density caused by the translation of the dipole-bound magnetic charges, expressed as the partial derivative of the magnetization field. Again, the behavior of the system is completely described by classical equations of physics in Table 2. No external forces are needed and the total mechanical and electrodynamic energy in the system is conserved.

Next, the behavior of two electric current dipoles illustrated in Figure 7b) is considered. First, the behavior of two interacting electric current loops described by the classical laws of physics in Table 1 is explained. In other word, their behavior without bringing in external forces and energies is explained (comparable to two interacting, superconducting loops). To start, internal forces in dipole 2 ($d_2$) when moved in the direction of $\mathbf{x}_{V2}$ are considered. The aim is to illustrate the origin of inductive forces within the current loop. The Lorentz-force on $d_2$ is given by:

$$\mathbf{f}_{e,L1} = \mathbf{j}_{e,V2} \times \mathbf{b}_{V1} = (\nabla \times \mathbf{m}) \times \mathbf{b}_{V1}. \quad (85)$$

For better illustration of occurring phenomena, the electric current loop as illustrated in Figure 8 is considered. When moved in direction of $\mathbf{x}_{V2}$, next to (85) a second Lorentz-force component $\mathbf{f}_{e,L2}$ is generated by the movement of the loop itself. In Figure 8 this force on the positive charge density is illustrated (blue):

$$\mathbf{f}_{e,L2,+} = \rho_{e,V2,+} \frac{d\mathbf{x}_{V2}}{dt} \times \mathbf{b}_{V1} \\ \mathbf{f}_{e,L2,-} = \rho_{e,V2,-} \frac{d\mathbf{x}_{V2}}{dt} \times \mathbf{b}_{V1}. \quad (86)$$

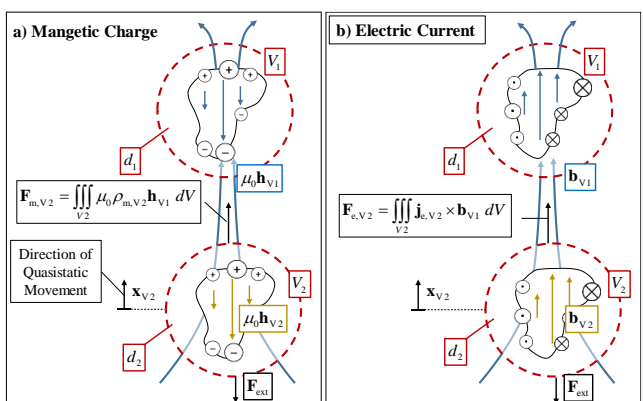

Figure 7: a) Illustration of two interacting electric current dipoles b) illustration of two interacting magnetic charge models.

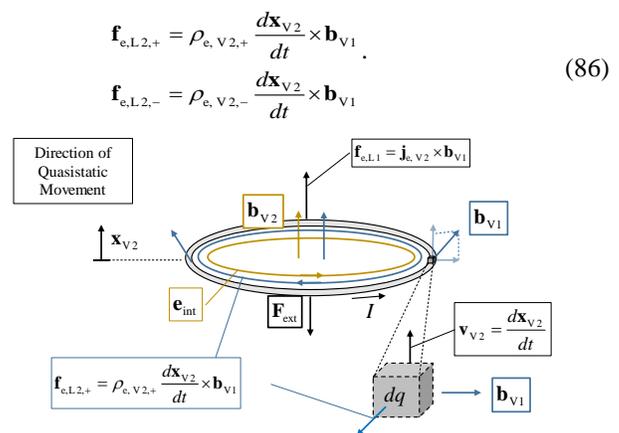

Figure 8: Schematic illustration of the Lorentz force (blue) and of the self-induced electric field (yellow) in a moving electric current loop.

This component of the total Lorentz force acts in the direction of current flow and must again be counterbalanced via a self-induced electric field by a change of the electric current in the loop. For the dipole in motion, the equivalent of (71) is:

$$\sum \mathbf{f}_{\rho_+} = 0 = \rho_{e,\,V2,+} \frac{d\mathbf{x}_{V2}}{dt} \times \mathbf{b}_{V1} + \rho_+ \mathbf{e}_{int} = \mathbf{f}_{e,L2,+} + \mathbf{f}_{e,+,int}$$
$$\sum \mathbf{f}_{\rho_-} = 0 = \rho_{e,\,V2,-} \frac{d\mathbf{x}_{V2}}{dt} \times \mathbf{b}_{V1} + \rho_- \mathbf{e}_{int} = \mathbf{f}_{e,L2,-} + \mathbf{f}_{e,-,int} \qquad (87)$$

The force which changes the electric current of the moving dipole is thus not a Coulomb force as for the stationary dipole (section III, C), but a component of the Lorentz force induced by the movement of the loop itself. The total Lorentz force (sum of (85) and (86)) acts perpendicular to the total charge velocity (sum of internal loop charge velocity and the external movement of the loop) and thus does no work. Instead, the electrical power from the counter inductive electric field is converted into mechanical power. The Lorentz force redirects the momentum and thereby the power flow by 90 degrees.

Next, it is considered what happens in terms of energy for the two interacting, superconducting loops. As the dipole is moved into the direction of force, mechanical work is done and the electromagnetic energy of the system

$$U_{e,em} = \int \frac{1}{2\mu_0} \mathbf{b} \cdot \mathbf{b} dV = \int \frac{1}{2} \mathbf{m} \cdot \mathbf{b} dV \cdot \qquad (88)$$

decreases. The total electromechanical energy of the system is conserved. However, the electric currents in both dipoles are not staying constant. The total field energy is considered in more detail by separating the fields of the individual dipole terms:

$$\mathbf{b} = \mathbf{b}_{V1} + \mathbf{b}_{V2} \qquad \mathbf{m} = \mathbf{m}_{V1} + \mathbf{m}_{V2} \cdot \qquad (89)$$

The total energy may be divided into the mutual and own energy terms of the two dipoles:

$$U_{e,em} = \int \frac{1}{2\mu_0} \mathbf{b}_{V1} \cdot \mathbf{b}_{V1} dV + \int \frac{1}{2\mu_0} \mathbf{b}_{V2} \cdot \mathbf{b}_{V2} dV$$
$$+ \int \frac{1}{\mu_0} \mathbf{b}_{V1} \cdot \mathbf{b}_{V2} dV \qquad (90)$$

As the currents and corresponding fields $\mathbf{b}_{V1}$ and $\mathbf{b}_{V2}$ of both dipoles decrease when moved in the direction of force, the individual energy terms of both dipoles decrease as well:

$$\frac{\partial \left( \int \frac{1}{2\mu_0} \mathbf{b}_{V1} \cdot \mathbf{b}_{V1} dV \right)}{\partial \mathbf{x}_{V2}} = \frac{\partial \left( \int \frac{1}{2\mu_0} \mathbf{b}_{V2} \cdot \mathbf{b}_{V2} dV \right)}{\partial \mathbf{x}_{V2}} < 0 \cdot \qquad (91)$$

In opposition to the magnetic charge model, the mutual energy term of the electric current model increases:

$$\frac{\partial \left( \int \frac{1}{\mu_0} \mathbf{b}_{V1} \cdot \mathbf{b}_{V2} dV \right)}{\partial \mathbf{x}_{V2}} = \frac{\partial \left( \int \mathbf{m}_{V1} \cdot \mathbf{b}_{V2} dV \right)}{\partial \mathbf{x}_{V2}} > 0 \cdot \qquad (92)$$

The different signs in the derivatives of the mutual energy term of the two models may be explained by considering the fields inside the dipoles: the fields $\mathbf{b}_{V1}$ and $\mathbf{b}_{V2}$ inside the electric current dipoles point in the same directions, the fields $\mathbf{h}_{V1}$ and $\mathbf{h}_{V2}$ point in opposite directions. As no external force or energy is brought in the system, the electromechanical energy is conserved, and the total energy decreases according to the mechanical work done. The force equals the change in all field energies:

$$\mathbf{F}_{e,V2} = \int \mathbf{j}_{V2} \times \mathbf{B}_1 dV$$
$$= -\frac{\partial U_{em,e}}{\partial \mathbf{x}_2} = -\left( \frac{\partial U_{em,e1}}{\partial \mathbf{x}_2} + \frac{\partial U_{em,e2}}{\partial \mathbf{x}_2} + \frac{\partial U_{em,e,mut}}{\partial \mathbf{x}_2} \right) \cdot \qquad (93)$$

The system of two interacting superconducting electric current dipoles, described by the classical laws of physics in Table 1, behaves differently than two interacting magnetic charge dipoles described by the laws in Table 2. The electric currents and the corresponding vector fields $\mathbf{m}_{V1}$ and $\mathbf{m}_{V2}$ of the electric current dipole change when moved. As the currents change, also the corresponding forces on the dipoles change. This is not the behavior aimed to model for the electron-spin, as the spin is presupposed to be a constant property.

To keep the electric current in dipole 2 constant, the induced force by the movement of the dipole needs to be compensated externally to obviate the need for the self-induce electric field:

$$\mathbf{f}_{\rho_+,ext} = -\rho_{e,\,V2,+} \frac{d\mathbf{x}_2}{dt} \times \mathbf{b}_{V1} \cdot$$
$$\mathbf{f}_{\rho_-,ext} = -\rho_{e,\,V2,-} \frac{d\mathbf{x}_2}{dt} \times \mathbf{b}_{V1} \qquad (94)$$

This force is also called motional electromotive force and is of the same magnitude as the Coulomb force in (71) [4, p. 462]. Following, the external energy needed, to keep the current constant is the same as in the non-moving dipole ((76)):

$$U_{ext,\,2} = -\int \mathbf{m} \cdot \mathbf{b}_{ext} dV \cdot \qquad (95)$$

Again, the strange behavior of the external force may be noted. For the moving loop, the external force compensates a part of the Lorentz force, in the stationary loop it compensates the induced Coulomb force. (Side note: I see no simple way to extend the basic equations of classical physics in Table 1 and mathematically express the described external forces that are required to keep the electric current dipole constant.)

The consequences of bringing in external, non-electromagnetic forces may be illustrated when considering the change of the total electromagnetic energy of the system. As one dipole is moved in the direction of force, mechanical work is done on the system. As the currents of the two dipoles stay constant, the electromagnetic energies of the individual dipoles $U_{e,em,1}$ and $U_{e,em,2}$ stay constant:

$$\frac{\partial \left( \int \frac{1}{2\mu_0} \mathbf{b}_{V1} \cdot \mathbf{b}_{V1} dV \right)}{\partial \mathbf{x}_{V2}} = \frac{\partial \left( \int \frac{1}{2\mu_0} \mathbf{b}_{V2} \cdot \mathbf{b}_{V2} dV \right)}{\partial \mathbf{x}_{V2}} = 0 \cdot \qquad (96)$$

As shown before, the mutual energy increases when moved in the direction of force:

$$\frac{\partial \left( \int \frac{1}{\mu_0} \mathbf{b}_{V1} \cdot \mathbf{b}_{V2} dV \right)}{\partial \mathbf{x}_{V2}} = \frac{\partial \left( \int \mathbf{m}_{V1} \cdot \mathbf{b}_{V2} dV \right)}{\partial \mathbf{x}_{V2}} > 0 \cdot \qquad (97)$$

The total mechanic and electrodynamic energies in the system, following from the laws of classical physics in Table 1 are not conserved. When mechanical work is done, the electromagnetic field energy of the system increases. This is the consequences of bringing in external forces and energies into the system, to keep the electric current model constant. The classical laws of physics do not represent the behavior of the constant electric current dipole.

When aiming to calculate the force on a dipole by variation of energy, the external energies need to be considered. The external energies, to keep two dipoles constant, are given by (76) and (95):

$$U_{ext} = -\int \mathbf{m}_{V1} \cdot \mathbf{b}_{V2} dV - \int \mathbf{m}_{V2} \cdot \mathbf{b}_{V1} dV \cdot \tag{98}$$

The force on dipole 2 (Figure 7b)) is calculated by the variation of all energies:

$$
\begin{aligned}
\mathbf{F}_{e,V2} &= \int \mathbf{j}_{V2} \times \mathbf{b}_1 dV = -\frac{\partial \left( U_{em,e} + U_{ext} \right)}{\partial \mathbf{x}_2} \\
&= -\frac{\partial \left( U_{em,e1} + U_{em,e2} + U_{em,e,mut} + U_{ext} \right)}{\partial \mathbf{x}_2} \\
&= -\frac{\partial \left( U_{em,e,mut} + U_{ext} \right)}{\partial \mathbf{x}_2} \\
&= -\frac{\partial \int \mathbf{m}_{V1} \cdot \mathbf{b}_{V2} dV - 2\int \mathbf{m}_{V1} \cdot \mathbf{b}_{V2} dV}{\partial \mathbf{x}_2} \\
&= -\frac{\partial \int -\mathbf{m}_{V1} \cdot \mathbf{b}_{V2} dV}{\partial \mathbf{x}_2}
\end{aligned}
\tag{99}
$$

The result is the same force as in the system with magnetic charges (84), as already shown in chapter III B.

In summary, it was shown that the constant electric current model is not a closed electromechanical system. The constant electron spin cannot be described by the fundamental laws of classical physics, which are given in Table 1. It needs external forces and external energies to keep the electric current constant. The consequence of bringing in external energies, is that the electrodynamic and mechanic energies of the system are not conserved. The dipole model consisting of magnetic charges avoids all these problems. It is a closed electromechanical system, it does not need external forces or energies and it is completely described by the laws of classical physics given in Table 2.

## IV. SIZE AND DISTANCES OF INTERACTING ELECTRONS IN FERROMAGNETIC MATTER

In chapter III the behavior of bound electric and magnetic sources as representations of the electron spin was compared. So far, no restrictions about the form or the size of the electron and the corresponding vector field $\mathbf{m}$ were made. In this work, solid ferromagnetic matter in mechatronic systems is considered. The matter is considered to be operating at temperatures at which the atoms are arranged in a crystalline lattice as illustrate in Figure 1. The distance between neighboring electrons is given by the lattice constant of a material. Examples of lattice constants of materials used in mechatronic systems are:

- Iron : $\sim 280$ pm $= 2.8 \cdot 10^{(-10)}$ m
- Nickel : $\sim 350$ pm $= 3.5 \cdot 10^{(-10)}$ m
- Cobalt : $\sim 250$ pm $= 2.5 \cdot 10^{(-10)}$ m
- Neodymium- Iron- Boron : $\sim 400$ pm $= 4 \cdot 10^{(-10)}$ m

The concept of size of an electron in classical physics is not straightforward as it is better understood within the framework of quantum mechanics for which the electrons are described by wave functions that give probabilities for finding the electron at different locations. However, in classical electrodynamics, for dimensions large compared to $10^{-14}$ m , the electron may be regarded as a point particle, meaning it has no size in the traditional sense [2, p. 248]. The goal in this section is to reconstruct this idea of a point-like dipole without giving up the understanding of the electron as a distributed system. Thereby the same starting point for the derivation of the macroscopic Maxwell equations as in the regarded books on classical electrodynamics is obtained.

The vector potential of the magnetic field of an arbitrary shaped electric current dipole may be divided into different parts by a multipole expansion [4, p. 336]. The first non-zero term is given by:

$$A_k(\mathbf{x}) = \frac{\mu_0}{4\pi} \frac{\mathbf{m}_p \times (\mathbf{x} - \mathbf{x}_0)}{|\mathbf{x} - \mathbf{x}_0|^3} \cdot \tag{100}$$

where $\mathbf{m}_p$ is defined to be the dipole moment of the current distribution:

$$\mathbf{m}_p = \frac{1}{2} \int \mathbf{j}_{e,b}(\mathbf{x}) \times \mathbf{x} dV = \frac{1}{2} \int (\nabla \times \mathbf{m}(\mathbf{x})) \times \mathbf{x} dV \cdot \tag{101}$$

The integral reduces the distributed system $\mathbf{j}_{e,b}(\mathbf{x})$ or $\mathbf{m}(\mathbf{x})$ to a point system $\mathbf{m}_p$ located at $\mathbf{x}_0$. The position $\mathbf{x}_0$ does not result out of the integral but needs to be assigned to $\mathbf{m}_p$ externally (e.g. in the center of mass of current distribution). As the electron size is assumed to be negligible small compared to the distance of interacting electrons, the remaining terms of the multipole expansion of the vector potential of an arbitrary current distribution are negligible. All relevant, interacting fields may be described by the concentrated parameter $\mathbf{m}_p$, as the exact current distribution within the volume is irrelevant for the interacting fields. The magnetic field of the magnetic dipole moment $\mathbf{m}_p$ is given by [12, 8, 4, p. 343]:

$$\mathbf{b}_p(\mathbf{x}) = \frac{\mu_0}{4\pi} \left[ \frac{3\mathbf{n}(\mathbf{m}_p \cdot \mathbf{n}) - \mathbf{m}_p}{|\mathbf{x} - \mathbf{x}_0|^3} \right] + \frac{2\mu_0 \mathbf{m}_p}{3} \delta(\mathbf{x} - \mathbf{x}_0) \cdot \tag{102}$$

Next, the change of the momentum of the electron is considered, which is given by:

$$\frac{d\mathbf{P}_{e,cm}}{dt} = \int \mathbf{j}_{e,b} \times \mathbf{b}_{ext} \ dV + \frac{d\int \varepsilon_0 \mathbf{e}_{ext} \times \mathbf{b} \ dV}{dt} \cdot \tag{103}$$

The external fields are approximated by a Taylor series at the location of the regarded electron $\mathbf{x}_0$:

$$\mathbf{b}(\mathbf{x}) = \mathbf{b}(\mathbf{x}_0) + \nabla_p \mathbf{b}(\mathbf{x}_0)(\mathbf{x} - \mathbf{x}_0) + O^2 \cdot \tag{104}$$

The first term $\mathbf{b}(\mathbf{x}_0)$ of (104) in (103) integrates to zero. As the volume is negligibly small compared to the size of the system of interacting electrons, the higher order force terms resulting from $O^2$ in (104) are negligible. (103) may be reformulated to (as shown in [18] [14])):

$$\begin{aligned} \frac{d\mathbf{P}_{e,cm}}{dt} &= \nabla\left(\mathbf{m}_p \cdot \mathbf{b}_{ext}(\mathbf{x}_0)\right) + \frac{d\mu_0\varepsilon_0\mathbf{e}_{ext}(\mathbf{x}_0)\times\mathbf{m}_p}{dt} \\ &= \left(\mathbf{m}_p \cdot \nabla\right)\mathbf{b}_{ext}(\mathbf{x}_0) + \mathbf{m}_p \cdot \left(\nabla\times\mathbf{b}_{ext}(\mathbf{x}_0)\right) \\ &\quad + \frac{d\mu_0\varepsilon_0\mathbf{e}_{ext}(\mathbf{x}_0)\times\mathbf{m}_p}{dt} \end{aligned} \tag{105}$$

The change of the linear momentum of the center of mass of the electron is only dependent on $\mathbf{m}_p$. The exact current distribution in the volume is irrelevant if the electron size is considered to be negligibly small compared to the size of the system of interacting electrons.

Due to the implied condition on size and distances of interacting electrons, the force and the interacting fields of the electrons are described completely by the point entity $\mathbf{m}_p$. The electron model has been reduced from a distributed system with current density $\mathbf{j}_e(\mathbf{x})$ to a single point $\mathbf{m}_p$ located at an assigned location $\mathbf{x}_0$. Although the electron is still considered to be a distributed system, all relevant phenomena are adequately described by $\mathbf{m}_p$. The magnetic field (102) has a corresponding "theoretical" electric current distribution [4, p. 343]:

$$\mathbf{j}_{e,b}(\mathbf{x}) = \nabla\times\mathbf{m}_p\delta(\mathbf{x}-\mathbf{x}_0) \cdot \tag{106}$$

with the corresponding magnetization field:

$$\mathbf{m}(\mathbf{x}) = \mathbf{m}_p\delta(\mathbf{x}-\mathbf{x}_0) \cdot \tag{107}$$

This object is called an ideal electric current dipole. The external power $P_{ext}$ needed to keep an electric current dipole constant was derived to be:

$$P_{ext} = -\int\mathbf{m}(\mathbf{x})\mathbf{b}_{ext}dV \cdot \tag{108}$$

Inserting (107) into (108), for the point-like ideal electric current dipole it follows:

$$P_{ext} = -\mathbf{m}_p\mathbf{b}_{ext}(\mathbf{x}_0) \cdot \tag{109}$$

The external energy to keep the ideal electric current constant is dependent on $\mathbf{m}_p$. An ideal dipole is a theoretical model of which the corresponding electric current density (106) approaches infinity and the size of the dipole approaches zero [5, p. 155]. It is emphasized that the electron is not considered to be infinitely small, as this would lead its energy to be infinite. Inserting (102) into (10) yields:

$$\begin{aligned} U_{e,em} &= \int\frac{1}{2\mu_0}\mathbf{b}_p \cdot \mathbf{b}_p dV \\ &= \frac{1}{2}\int...\delta(\mathbf{x}-\mathbf{x}_0)^2 dV = \infty \end{aligned} \tag{110}$$

The analog problem arises when treating an electron as a point charge as shown in [1, p. 192]. It is aimed to avoid this problem and assign the dipole a small distribution while the size of the dipole is negligible small compared to the regarded system. Everything presented is this chapter is transferable to the magnetic charge dipole. The field of the ideal magnetic charge dipole is given by:

$$\begin{aligned} \mathbf{h}_p(\mathbf{x}) &= \frac{1}{\mu_0}\mathbf{b}_p(\mathbf{x}) - \mathbf{m}_p\delta(\mathbf{x}-\mathbf{x}_0) \\ &= \frac{1}{4\pi}\left[\frac{3\mathbf{n}(\mathbf{m}_p\cdot\mathbf{n}) - \mathbf{m}_p}{|\mathbf{x}-\mathbf{x}_0|^3}\right] - \frac{1}{3}\mathbf{m}_p\delta(\mathbf{x}-\mathbf{x}_0) \end{aligned} \tag{111}$$

The change of momentum of the ideal magnetic dipole is given by [14]:

$$\frac{d\mathbf{P}_{m,cm}}{dt} = \mu_0\left(\mathbf{m}_p\cdot\nabla\right)\mathbf{h}_{ext}(\mathbf{x}_0) + \mu_0\varepsilon_0\frac{d\mathbf{m}_p}{dt}\times\mathbf{e}_{ext}(\mathbf{x}_0) \cdot \tag{112}$$

As no external sources are within the dipole volume the change of momentum of both models ((105) and (112)) equal, as shown in [14].

This chapter was introduced to illustrate that under given assumptions about size and distances of the interacting electrons, the exact distribution of the currents or charges inside the volumes are not relevant. Thereby, the point-wise dipole representation employed in common textbooks is correlated to the model of an arbitrary shaped dipole. This leaves the same starting point for the derivation of the macroscopic Maxwell equation in following work.

## V. Comment on Potential magnetic Energy (Zeeman Energy)

Before summarizing and discussing the findings of this paper, a short discussion on the meaning and derivation of the potential magnetic energy is conducted. Therefore, a different approach for the inclusion of the electron spin in electrodynamics is considered, as suggested in [5, p. 378]. The idea is to incorporate the electron spin in Maxwell's field and force equations and not associate it with a magnetic or electric source at all, but merely as a permanent magnetic point dipole $\mathbf{m}_p$. To try this, first, the dipole field $\mathbf{m}$ is included into Maxwell's field equations:

$$\begin{aligned} \nabla\cdot\mathbf{e} &= \rho_e \\ \nabla\times\mathbf{e} &= -\frac{\partial\mathbf{b}}{\partial t} \\ \nabla\cdot\mathbf{b} &= 0 \\ \frac{1}{\mu_0}\nabla\times\mathbf{b} &= \mathbf{j}_e + \nabla\times\mathbf{m} + \varepsilon_0\frac{\partial\mathbf{e}}{\partial t} \end{aligned} \tag{113}$$

$\mathbf{m}$ may be considered the point-like entity in (107), however, it is not important whether $\mathbf{m}$ is point-like or distributed for the point aimed to illustrate. As shown by (113) there is no difference between an electric current $\mathbf{j}_e$ and $\nabla\times\mathbf{m}$ in terms of fields. The force density on the permanent magnetic dipole is given by:

$$\mathbf{f}_{e,m} = \left(\nabla\times\mathbf{m}\right)\times\mathbf{b} \cdot \tag{114}$$

For a point dipole, the total force may be reformulated to ([4, pp. 373-374]):

$$\mathbf{F}_{e,m} = \nabla\left(\mathbf{m}_p\cdot\mathbf{b}\right) \cdot \tag{115}$$

So far, there is no difference between a free current and the permanent magnetic point dipole $\mathbf{m}_p$. Up to this point, there was equally no relevant difference between the magnetic charge model and the electric current model. It is crucial to consider carefully how the individual models generate power. For the constant magnetic dipoles, it may be assumed that the power is generated by:

$$P_{\mathrm{m}} = -\mathbf{F}_{\mathrm{c,em}} \cdot \mathbf{v} = -\nabla\left(\mathbf{m}_{\mathrm{p}} \cdot \mathbf{b}\right) \cdot \frac{d\mathbf{x}}{dt} = -\frac{d\left(\mathbf{m}_{\mathrm{p}} \cdot \mathbf{b}\right)}{dt} = \frac{dU_{\mathrm{m,pot}}}{dt} \ . \quad (116)$$

In this model, $\mathbf{v}$ is the velocity of the dipole and $\mathbf{F}_{\mathrm{c,em}}$ is a conservative force on the dipole with the corresponding potential energy $U_{\mathrm{m,pot}}$ (sometimes called Zeeman energy):

$$U_{\mathrm{m,pot}} = -\mathbf{m} \cdot \mathbf{b} \ . \quad (117)$$

However, a model which generates power by (116) (called potential energy model in this work) fundamentally differs from the described electric current and magnetic charge model. If it is assumed that the power is generated by (116), classical electrodynamics in terms of a field theory is abandoned. The potential energy $U_{\mathrm{m,pot}}$ is not localized in the field and thus in every point in space but is a property of the system as a whole. In this model, there are no fields that propagate through space and carry energy and momentum. This simplified model of interacting electron spins may be compared to the Newtonian gravitation model. There, gravity is not a field theory, but rather a theory of instantaneous action-at-a-distance, where objects exert gravitational forces on each other directly across space without any intermediary medium or field.

The energies of the potential energy model of two interacting dipoles before and after bringing them together is illustrated in Figure 9. When the two dipoles are infinitely distant, the total energy of the system following (116) is zero. There is no field energy $u_{\mathrm{e,em}} = \frac{1}{2\mu_0}\mathbf{b} \cdot \mathbf{b}$ that is derived from the given power equation (116). When bringing the dipoles together, the change in potential energy equals the mechanical energy $U_{\mathrm{mech}}$ as illustrated in Figure 9. This potential energy model may be used in very selected static cases to calculate forces. However, it does not represent the electromagnetic field theory and is a strong simplification of occurring phenomena.

If the goal is to hold the field theory, the power of the constant dipoles needs to equal:

$$P_{\mathrm{e,em}} = \nabla \cdot \mathbf{S}_{\mathrm{e}} + \frac{\partial u_{\mathrm{e,em}}}{\partial t} \ , \quad (118)$$

with the field energy

$$u_{\mathrm{e,em}} = \frac{\varepsilon_0}{2}\mathbf{e} \cdot \mathbf{e} + \frac{1}{2\mu_0}\mathbf{b} \cdot \mathbf{b} \ , \quad (119)$$

and the poynting vector representing the power flow:

$$\mathbf{S}_{\mathrm{e}} = \frac{1}{\mu_0}\mathbf{e} \times \mathbf{b} = c^2 \mathbf{g}_{\mathrm{e}} \ . \quad (120)$$

This behavior results when the constant dipole generates power like a free electric current:

$$P_{\mathrm{e,em}} = \left(\nabla \times \mathbf{m}\right) \cdot \mathbf{e} = \nabla \cdot \mathbf{S}_{\mathrm{e}} + \frac{\partial u_{\mathrm{e,em}}}{\partial t} \ . \quad (121)$$

By developing this model further, the resulting behavior aligns precisely with that of the constant electric current model described in the previous chapters: the power in (121) needs to correspond to a force density and velocity:

$$P_{\mathrm{e,em}} = \mathbf{f}_{\mathrm{e,em}} \cdot \mathbf{v} = \left(\nabla \times \mathbf{m}\right) \cdot \mathbf{e} \ . \quad (122)$$

Following (122), the force density must generate a force in an external electric field. However, it must still hold that $\nabla \times \mathbf{m}$ generates no total force in an electric field. The solution is to separate $\nabla \times \mathbf{m}$ into two terms of different velocity:

$$\left(\nabla \times \mathbf{m}\right) = \rho_{m,+} \cdot \mathbf{v}_{+} + \rho_{m,-} \cdot \mathbf{v}_{-} \ . \quad (123)$$

The development of this model is ended at this point, as it essentially reproduces the electric current loop model discussed in previous chapters. The aim was to incorporate a constant magnetic dipole into Maxwell's field and force equations without relating $\mathbf{m}_{\mathrm{p}}$ to any sources. However, to describe its behavior in terms of fields, the same model of an electric current loop was derived which was detailly described in in the last chapter.

To compare the electric current model to the potential energy model, the energies before and after bringing two electric current dipoles together are illustrated in Figure 10. When infinitely far away, the system's energies are given by the field energies of the individual dipoles $U_{\mathrm{e,em,1}}$ and $U_{\mathrm{e,em,2}}$. When brought together, the external energies $U_{\mathrm{ext,1}}$ and $U_{\mathrm{ext,2}}$ needed to keep the current constant, are brought into the system. These equal the sum of the change in field energy $U_{\mathrm{e,mutual}}$ and the mechanical work $U_{\mathrm{mech}}$. It is emphasized that the external

**Potential Energy Model**

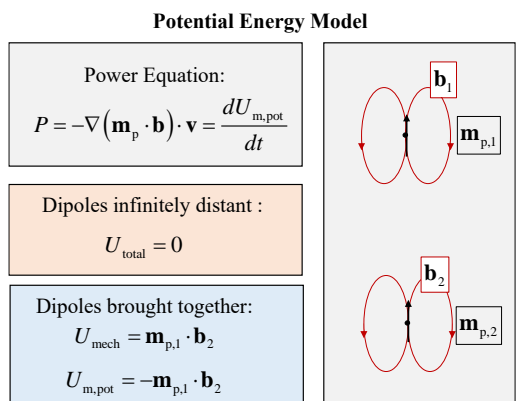

Figure 9: Power and energy of two interacting constant magnetic dipoles in the electric current model.

**Electric Current Model**

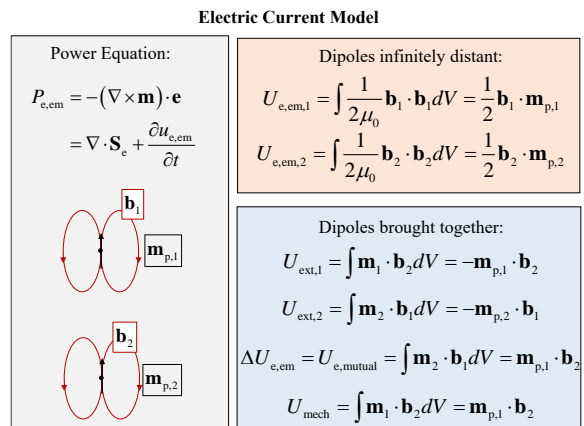

Figure 10: Power and energy of two interacting constant magnetic dipoles in the Potential Energy Model.

energies $U_{ext}$ should not be exchanged with the potential energy $U_{m,pot}$ from the potential energy model. There is no basis for including the potential energy as well as electromagnetic field energies $U_{e,em}$ in the same system as the two energies are associated with different power models and thus should be considered separately.

Figure 11 illustrates the same system of two interacting dipoles in the magnetic charge model. As explained previously, in this model everything is described by field energies and no external inputs are required. When infinitely far away, the system's energies are given by the field energies of the individual dipoles $U_{m,em,1}$ and $U_{m,em,2}$. When brought together, the mechanical work $U_{mech}$ equals the decrease in field energy $U_{m,mutual}$. The total field energy of the system is:

$$U_{m,em} = \int \frac{\mu_0}{2} \mathbf{h} \cdot \mathbf{h} \, dV \cdot \tag{124}$$

As shown in Appendix C, the total field energy might be expressed as:

$$U_{m,em} = -\int \frac{\mu_0}{2} \mathbf{m} \cdot \mathbf{h} \, dV \cdot \tag{125}$$

This term is sometimes referred to as a magnetic potential energy, however, it represents the total field energy of the system. The field energy may always be separated into two interacting systems:

$$U_{m,em} = U_{m,em,1} + U_{m,em,2} + U_{m,mutual} \tag{126}$$
$$= -\int \left( \frac{\mu_0}{2} \mathbf{m}_1 \cdot \mathbf{h}_1 - \frac{\mu_0}{2} \mathbf{m}_2 \cdot \mathbf{h}_2 - \mu_0 \mathbf{m}_1 \cdot \mathbf{h}_2 \right) dV \cdot$$

If the magnetization of the two interacting systems remains constant, the change in field energy is described by the change of the mutual energy:

$$U_{m,mutual} = -\int \mu_0 \mathbf{m}_1 \cdot \mathbf{h}_2 \, dV \tag{127}$$

Again, (127) should not be mistaken for the potential energy $U_{m,pot}$ from the potential energy model. In my opinion, the introduction of a potential energy model and the corresponding potential energy $U_{m,pot}$ leads to confusion and is not necessary for the description of magnetic dipoles. Everything should be described by field energies derived from the basic laws of classical electrodynamics.

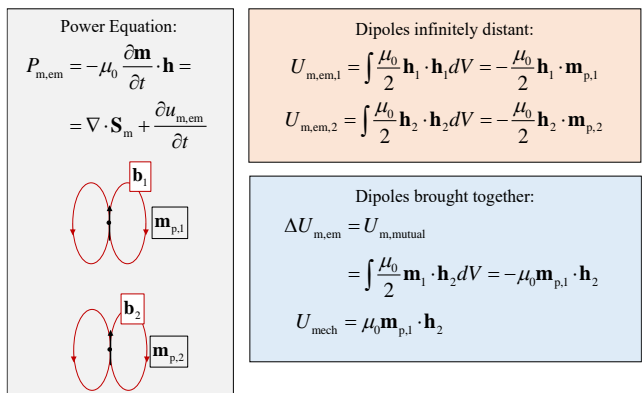

Figure 11: Power and energy of two interacting constant magnetic dipoles in the magnetic charge model.

## VI. Summary and Discussion

In this work, two classical electrodynamic models used to represent the magnetic dipole moment attributable to the quantum mechanical electron spin were compared. The electron spin was presupposed to be a constant, intrinsic property of the electron, also when interacting with external fields. It was shown that the constant electric current model of the electron spin is not a closed electromechanical system, as it needs external forces and energies which are not included in the fundamental laws of classical physics (Table 1). Consequently, the electromechanic energy in this model is not conserved. In opposition to this model, it was shown that the magnetic charge model of the electron is a closed electromechanical system, which may completely be described by the classical laws of physics in Table 2.

By investigating the phenomena occurring within the electron volume, it was illustrated that distinguishing between electric and magnetic sources as the representation of the constant magnetic dipole is infeasible solely through observations of the fields outside the dipole volume and the resultant total dipole force. Notably, differences in field distributions for a constant dipole emerge exclusively within the dipole volume, while discrepancies in total force necessitate an external source within this domain.

The presented example of hyperfine splitting has shown that for microscopic considerations in the dimensions of elementary particles (in the range of the Compton wavelength: electron $\sim 2.4 \cdot 10^{-12}$ m) phenomena occur that support the electric current model of the electron spin. To generate this phenomenon for an electron, the interaction of an electron and a positron in the exotic atom positronium was required while quantum field theory is needed to describe occurring effects.

However, when regarding the modelling of solid ferromagnetic matter for engineering applications in macroscopic mechatronic systems, such phenomena are too small to be observed or averaged out over large scales, making it reasonable to neglect their influence. In solid ferromagnetic matter, electrons are arranged in a crystal lattice and their distance is several orders of magnitude greater than their size. All relevant information is given by the point entity $\mathbf{m}_p$ while the exact continuum distribution of the sources of the electron spin model is negligible. The lattice arrangement of solid ferromagnetic matter and the point-like representation of the electron-spin imply that no electron (or other electric or magnetic source) resides inside another electron in this model. Quantum mechanical phenomena such as hyperfine splitting are not represented in this purely classical model. For modelling of ferromagnetic matter in mechatronic systems, these phenomena may be neglected as, to my knowledge, they have no relevant macroscopic effects on the systems.

Figure 12 schematically illustrates the differences of the microscopic magnetic fields of the two models of ferromagnetic matter. It shows that the field distributions solely differ inside the negligible small volumes of the point-like electrons. As no sources were assumed to be within the electron volumes, the behaviour and the momentum change of solid ferromagnetic matter cannot be distinguished for both models.

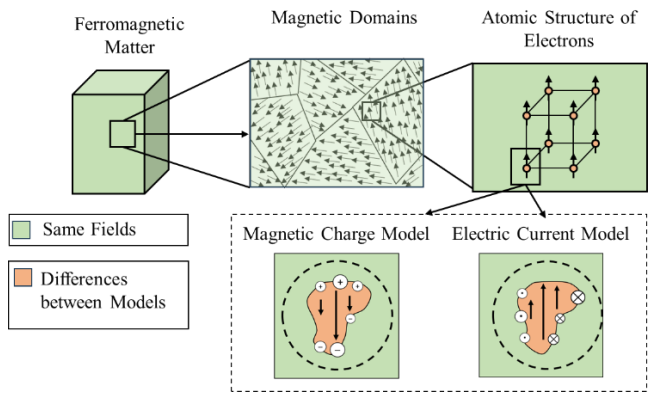

Figure 12: Illustration of the magnetic field differences for a microscopic description of the two dipole models.

The crucial difference between the two models became evident by examining the processes inside the electron volume in detail. It was demonstrated that the electric current model depends on external forces to maintain the force balance on the electric charge density inside the electron in order to keep the dipole constant. These external forces change dynamically and do work on the system as the electron interacts with external fields. Consequently, the model of a constant electric current dipole is not a closed electromechanical system. The energies derived from classical physics (gravitational potential energy, kinetic energy, electrodynamic field energy) are not conserved in the system. The model of a constant electric current cannot be described using the laws of classical physics in Table 1. Furthermore, it was shown that all mentioned problems may be circumvented by adding magnetic charges to the Maxwell equations and by modelling the electron spin as a magnetic charge dipole. The magnetic charge dipole model is a closed electromechanical system, it does not need external forces, the energies in the system are conserved, and it is completely described by the laws of classical physics given in Table 2.

## VII. FUTURE WORK

This work builds the basis for further considerations of ferromagnetic matter in mechatronic systems. In future work, the macroscopic Maxwell-Equations for both models will be derived, using averaging functions which rearrange the sources and corresponding fields. The macroscopic equations will be derived in two steps. First, an average function in a size range which merely averages the sources within the magnetic domains is used, as illustrated in Figure 13.

**Magnetic Domains**

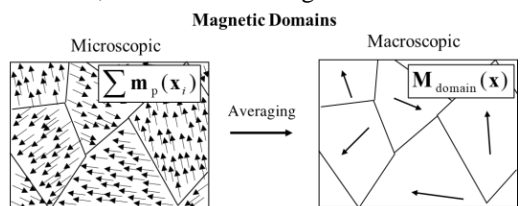

Figure 13: Microscopic and macroscopic magnetization field of the magnetic domains.

The consequences of the averaging process will be illustrated by pointing out under which conditions the microscopic and averaged macroscopic systems behave differently. It will be shown, how the domains align themselves internally and by interaction with external fields including the consideration of dissipative effects. The aim is to illustrate that the electric current model on the macroscopic level of interacting domains behaves exactly as the microscopic description of the electron spin presented in this paper. In other words, it will be shown that everything presented in this paper is exactly transferable to the macroscopic equations of interacting domains.

One exemplary consequence, which is often misleadingly presented in the literature, is that the domains align themselves towards the minimum electromagnetic energy [1, p. 903]:

$$U_{\text{e,em}} = \int \frac{1}{2\mu_0} \mathbf{B} \cdot \mathbf{B} dV \cdot \tag{128}$$

However, they align themselves towards the maximum electromagnetic energy $U_{\text{e,em}}$ within the electric current model but to the minimum electromagnetic energy

$$U_{\text{m,em}} = \int \frac{1}{2\mu_0} \mathbf{H} \cdot \mathbf{H} dV \tag{129}$$

within the magnetic charge model as illustrated in this work.

Moreover, the process of averaging from a domain level to a larger scale of ferromagnetic matter in mechatronic system will be considered in future work. This is schematically illustrated for a permanent magnet in Figure 14. The averaging process changes the system and rearranges the sources again. For permanent magnets, the magnetization field is considered constant, as the magnetic domains are regarded as fixed. All statements for the electron spin presented in this paper may be transferred and apply to permanent magnets on a macroscopic scale. However, for ferromagnetic matter in general, the volume bound magnetization of the model of macroscopic matter no longer remains constant as the orientation of the domains shifts when an external field is applied. In future work, the origin of the magnetic field energy

$$\iint \int \mathbf{H} \cdot d\mathbf{B} \, dV \cdot \tag{130}$$

which is derived by the externally supplied energy to magnetize ferromagnetic matter, will be described and compared for both models. The aim is to demonstrate that even at the macroscopic level, external energies and forces are required to describe the behavior of ferromagnetic matter using the electric current model.

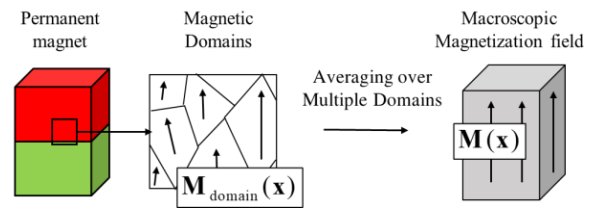

Figure 14: Microscopic and macroscopic magnetization field of the magnetic domains.

(Other future works will include the consideration of force density in ferromagnetic matter and the comparison of the two electron spin models in moving frames under consideration of relativistic effects. Moreover, the Lagrangian and Hamiltonian Methods may be considered in regard to this work.)

## VIII. Appendix

### A. Vector Field p – Corresponding Magnetic Sources

To illustrate how to exchange the electric sources corresponding to $\mathbf{p}$ by magnetic sources, first the variables in Table 2 are renamed:

$$\mathbf{d} = \varepsilon_0 \mathbf{e}, \quad \mathbf{b} = \mu_0 \mathbf{h} \,. \tag{131}$$

It follows:

$$\begin{aligned}
&\nabla \cdot \mathbf{d} = \rho_e \\
&-\frac{1}{\varepsilon_0}\nabla \times \mathbf{d} - \frac{\partial \mathbf{b}}{\partial t} = \mu_0 \mathbf{j}_m \\
&\nabla \cdot \mathbf{b} = \mu_0 \rho_m \\
&\frac{1}{\mu_0}\nabla \times \mathbf{b} - \frac{\partial \mathbf{d}}{\partial t} = \mathbf{j}_e
\end{aligned} \tag{132}$$

Next, (132), which is the system of merely electric sources, is reordered:

$$\begin{aligned}
&\nabla \cdot \left( \mathbf{e} + \frac{\mathbf{p}}{\varepsilon_0} \right) = 0 \\
&\nabla \times \mathbf{e} = -\frac{\partial \mathbf{b}}{\partial t} \\
&\nabla \cdot \mathbf{b} = 0 \\
&\frac{1}{\mu_0}\nabla \times \mathbf{b} = \frac{\partial (\varepsilon_0 \mathbf{e} + \mathbf{p})}{\partial t}
\end{aligned} \tag{133}$$

In foresight of what is aimed to derive, the new vector field $\mathbf{d}$ is introduced:

$$\mathbf{e} = \frac{1}{\varepsilon_0}(\mathbf{d} - \mathbf{p}) \quad \Leftrightarrow \quad \mathbf{d} = \varepsilon_0 \mathbf{e} + \mathbf{p} \,. \tag{134}$$

$\mathbf{d}$ has no physical significance regarding the system of electric sources. Inserting (134) into (133) yields:

$$\begin{aligned}
&\nabla \cdot \mathbf{d} = 0 \\
&\nabla \times \frac{\mathbf{d}}{\varepsilon_0} = -\frac{\partial \mathbf{b}}{\partial t} + \frac{1}{\varepsilon_0}\nabla \times \mathbf{p} \\
&\nabla \cdot \mathbf{b} = 0 \\
&\frac{1}{\mu_0}\nabla \times \mathbf{b} = \frac{\partial \mathbf{d}}{\partial t}
\end{aligned} \tag{135}$$

Comparing (135) with (132) it turns out that the source generating the fields $\mathbf{d}$ is the magnetic current density $\mathbf{j}_{m,b2}$:

$$\mathbf{j}_{m,b2} = c^2 \nabla \times \mathbf{p} \,. \tag{136}$$

The electric current density $\mathbf{j}_{e,b2}$ and electric charge density $\rho_{e,b}$ generated the same field outside the volume as the magnetic current density $\mathbf{j}_{m,b2}$, both represented by the vector field $\mathbf{p}$:

$$\rho_{e,b} = -\nabla \cdot \mathbf{p}, \; \mathbf{j}_{e,b2} = \frac{\partial \mathbf{p}}{\partial t} \; \leftrightarrow \; \mathbf{j}_{m,b2} = c^2 \nabla \times \mathbf{p} \,. \tag{137}$$

Figure 15 displays a stationary example of $\mathbf{p}$.

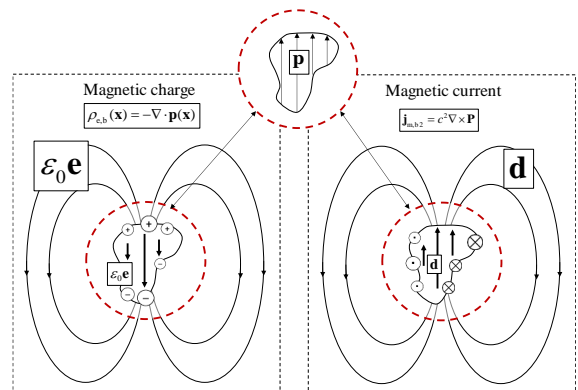

Figure 15: a) Electric charge representation of an arbitrary vector field p b) Magnetic current representation of an arbitrary vector field d

### B. Vector fields m and p are not uniquely defined

In chapter III A the vector fields $\mathbf{p}$ and $\mathbf{m}$ were introduced. It was remarked that $\mathbf{p}$ is not uniquely defined [4, p. 159]. Any solenoidal field $\nabla \times \mathbf{r}$, which holds the given boundary conditions, may be added to $\mathbf{p}$ without changing the associated charge distribution:

$$\rho_{e,b} = -\nabla \cdot \mathbf{p} = -\nabla \cdot (\tilde{\mathbf{p}} + \nabla \times \mathbf{r}) \,. \tag{138}$$

Moreover, it was remarked that $\mathbf{m}$ is not uniquely defined. Any irrotational field $\nabla \varphi_l$, which holds the given boundary conditions, may be added to $\mathbf{m}$ without changing the associated current distribution:

$$\mathbf{j}_{e,b} = \nabla \times \mathbf{m} = \nabla \times (\mathbf{m} + \nabla \varphi_l) \,. \tag{139}$$

There are certain vector fields $\mathbf{m}$ and $\mathbf{p}$ that only have a corresponding source in one of the two interpretations. In Figure 16a) a toroidal electric coil is illustrated for which there exists no corresponding magnetic charge distribution. The corresponding magnetization field is a solenoidal field. The pendant of an irrational magnetization field is illustrated in Figure 16b). For a charged hollow sphere surrounded by a neutralizing charged hollow sphere, there exists no corresponding source on the electric current side. This declares in which way a change in $\mathbf{m}$ or $\mathbf{p}$ may only have an impact in one of the two corresponding sources. It should be emphasized that in these cases no fields are generated outside the dipole volume, thus the fields still differ only within the volume. Also the magnetic dipole moments $\mathbf{m}_p$ ((101)) of these source are zero.

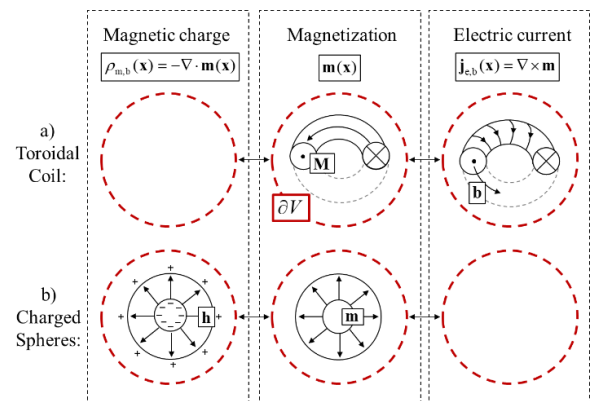

Figure 16: Magnetization field and the corresponding magnetic and electric source of a) an electric toroidal coil and b) neutralizing charged magnetic spheres.

### C. Magnetic Field Energy in terms of m

The goal is to show that for a stationary electric current dipole, it holds:

$$U_{e,em} = \int \frac{1}{2\mu_0} \mathbf{b} \cdot \mathbf{b} dV = \int \frac{1}{2} \mathbf{m} \cdot \mathbf{b} dV \cdot \qquad (140)$$

First

$$\mathbf{b} = \mu_0 \left( \mathbf{h} + \mathbf{m} \right) \qquad (141)$$

is inserted in (140). It follows:

$$\int \frac{1}{2\mu_0} \mathbf{b} \cdot \mathbf{b} dV = \int \frac{1}{2} \mathbf{h} \cdot \mathbf{b} dV + \int \frac{1}{2} \mathbf{m} \cdot \mathbf{b} dV \cdot \qquad (142)$$

Following the Maxwell equations in (29), for the stationary electric current dipole, it holds:

$$\nabla \times \mathbf{h} = 0 \cdot \qquad (143)$$

$$\nabla \cdot \mathbf{b} = 0 \cdot \qquad (144)$$

For irrotational and solenoid vector fields, it holds that:

$$\int \mathbf{h} \cdot \mathbf{b} dV = 0 \cdot \qquad (145)$$

It follows:

$$\int \frac{1}{2\mu_0} \mathbf{b} \cdot \mathbf{b} dV = \int \frac{1}{2} \mathbf{m} \cdot \mathbf{b} dV \cdot \qquad (146)$$

Magnetic Charge Model:

The goal is to show that for a magnetic charge dipole, it holds:

$$U_{e,em} = \int \frac{\mu_0}{2} \mathbf{h} \cdot \mathbf{h} dV = -\int \frac{\mu_0}{2} \mathbf{m} \cdot \mathbf{h} dV \cdot \qquad (147)$$

First

$$\mathbf{h} = \frac{1}{\mu_0} \mathbf{b} - \mathbf{m} \qquad (148)$$

is inserted. It follows:

$$\int \frac{\mu_0}{2} \mathbf{h} \cdot \mathbf{h} dV = \int \frac{1}{2} \mathbf{b} \cdot \mathbf{h} dV - \int \frac{\mu_0}{2} \mathbf{m} \cdot \mathbf{h} dV \cdot \qquad (149)$$

Following the Maxwell equations in (32), for the stationary magnetic charge dipole, it holds:

$$\nabla \times \mathbf{h} = 0 \cdot \qquad (150)$$

$$\nabla \cdot \mathbf{b} = 0 \cdot \qquad (151)$$

For irrotational and solenoid vector fields, it holds that:

$$\int \mathbf{h} \cdot \mathbf{b} dV = 0 \cdot \qquad (152)$$

It follows:

$$U_{e,em} = \int \frac{\mu_0}{2} \mathbf{h} \cdot \mathbf{h} dV = -\int \frac{\mu_0}{2} \mathbf{m} \cdot \mathbf{h} dV \cdot \qquad (153)$$